# Evidence of defect formation in monolayer MoS$_2$ at ultralow accelerating voltage electron irradiation

*Ajit Kumar Dash[1], Hariharan Swaminathan[1], Ethan Berger[2], Mainak Mondal[1], Touko Lehenkari[2], Pushp Raj Prasad[1], Kenji Watanabe[3], Takashi Taniguchi[4], Hannu-Pekka Komsa[2], Akshay Singh[1, \*]*

[1]Department of Physics, Indian Institute of Science, Bengaluru, Karnataka -560012, India

[2]Microelectronics Research Unit, University of Oulu, FI-90014, Oulu, Finland

[3]Research Center for Functional Materials, National Institute for Materials Science, Ibaraki 305-0044, Japan

[4]International Center for Materials Nanoarchitectonics, National Institute for Materials Science, Ibaraki 305-0044, Japan

*Corresponding author: aksy@iisc.ac.in

ABSTRACT:

Control on spatial location and density of defects in 2D materials can be achieved using electron beam irradiation. Conversely, ultralow accelerating voltages ($\leq$ 5kV) are used to measure surface morphology, with no expected defect creation. We find clear signatures of defect creation in monolayer (ML) MoS$_2$ at these voltages. Evolution of E' and A$_1$' Raman modes with electron dose, and appearance of defect activated peaks indicate defect formation. To simulate Raman spectra of $MoS_2$ at realistic defect distributions, while retaining density-functional theory accuracy, we combine machine-learning force fields for phonons and eigenmode projection approach for Raman tensors. Simulated spectra agree with experiments, with sulphur vacancies as suggested defects. We decouple defects, doping and carbonaceous contamination using control (hBN covered and encapsulated $MoS_2$) samples. We observe cryogenic PL quenching and defect peaks, and find that carbonaceous contamination does not affect defect creation. These studies have applications in photonics and quantum emitters.

**Introduction:**

Atomically thin transition metal dichalcogenides (TMDs) have exciting optoelectronic properties, including direct bandgaps and tightly bound excitonic complexes[1–4]. Defect engineering can further tune optoelectronic properties[5], and enable creation of novel functionalities[6]. In 2D TMDs, defects can be of zero (vacancies, interstitial atoms, antisites) or one-dimensional (dislocations, line defects) type[7]. Specifically, antisites and point defects in 2D TMDs provide an ideal platform for making spin qubits[8] and single-photon emitters (SPEs)[9–13] respectively. Defects can be unintentionally created during mechanical and chemical exfoliation, or chemical synthesis. Alternatively, defects can be deliberately introduced and spatially controlled by electron (or ion) beam irradiation.

Electron beam (e-beam) transfers energy to the sample by elastic and inelastic scattering mechanisms, including sputtering ($A_{sput}$) and radiolysis (R), and can also lead to carbonaceous contamination (C) (see Fig. 1(a)). E-beam of energy above knock-on threshold (~80 kV for S atoms) can create chalcogen vacancies by sputtering[7,14,15]. Focused ion beam (FIB) can also create defects by atomic sputtering[16–18]. However, additional material removal can occur due to ion beam bombardment. Hence, FIB is not preferable for controlled defect formation, especially for creating SPEs that require low defect density. Thus, ultralow e-beam accelerating voltages (1 - 5 kV) need to be explored for controllably creating defects, while minimizing surrounding lattice damage, resulting in superior SPEs.

Defects are not expected to be created below knock-on voltages, and especially at ultralow electron accelerating voltages. In contrast, there are recent reports of defect formation at low voltages. One proposed mechanism is defect creation due to localization of electronic excitations to emerging defect sites, which reduces displacement threshold energy of sputtering atoms[19]. In another study, 1 kV electron irradiation followed by annealing resulted in nanopores, attributed to carbonaceous contaminants-induced lowering of activation energy of defect creation[20]. Contrary to this observation, Parkin et al.[15] reported absence of interaction of carbonaceous contaminants with the sample. Further, Yagodkin et al.[21] recently showed formation of charge-transfer excitons between carbonaceous contaminants and the $MoS_2$ sample. SPE formation will benefit greatly from understanding of defect formation at ultralow accelerating voltages and effects of carbonaceous contamination[22,23].

Raman and photoluminescence (PL) spectroscopy are widely used for determining layer number and material quality. In ML $MoS_2$, first-order Raman modes are denoted as E' and $A_1$' (instead of $E^1_{2g}$ and $A_{1g}$ due to altered crystal symmetry in ML)[15,24,25]. Typical PL spectrum of ML $MoS_2$ at room temperature shows A and B excitons at 1.85 eV and 2 eV respectively, with the A peak comprising of neutral excitons ($X^0$) and negatively charged trions ($X^-$) [4]. With increasing defect concentration in ML $MoS_2$, Raman E' peak shifts towards lower wavenumber (k), and full width half maximum (FWHM, $\Gamma$) increases[15,18]. However, $A_1$' peak is not significantly affected by mono-sulphur vacancies. Defects can activate ZO and LO peaks, on the left and right shoulder of E' and $A_1$' peaks respectively[16,18,26]. For high defect densities, LA-band is observed[20]. Upon electron doping, due to strong electron-phonon coupling, $A_1$' peak shifts to lower wavenumber with increased $\Gamma$ [27,28]. However, E' peak is not significantly affected by electron doping. Defects also result in significant PL quenching of $X^0$ and B excitons[29], with increase in low-energy defect-bound excitons at cryogenic temperatures.

Quantifying defect types and concentrations based on experimental Raman spectra is challenging, and will greatly benefit from comparison to Raman spectra obtained from density functional theory (DFT). Unfortunately, modelling systems with low concentration of randomly distributed defects require very large supercells, making calculation of phonons and Raman tensors via conventional computational methods impossible. To reduce computational cost of Raman tensors, Hashemi et al.[30] recently developed a method (denoted RGDOS) based on projection of large supercell phonons onto those of the unit cell, and it has already been applied to $Mo_xW_{1-x}S_2$ and $ZrS_xSe_{1-x}$ alloys[30,31], defects in $MoS_2$[32], and SnS[33]. There also exist methods to accelerate the calculation of phonons, for example using classical potential[34] or machine learning force fields (MLFF)[34,35]. Using projection of vibrations with MLFF lead to a very efficient scheme to study large supercells containing small density of defects.

We investigate the creation of defects in mechanically exfoliated ML $MoS_2$ using e-beam irradiation at ultralow accelerating voltages ($\leq$ 5 kV). We employ Raman and PL to characterize and quantify defects, without creating additional defects. Evolution of E' peak with electron dose is attributed to defect formation, whereas $A_1'$ peak changes are related to interplay of defect formation, doping due to oxygen adsorbates (physisorption and chemisorption) and e-beam. To understand defect types and concentrations, we combine MLFF and RGDOS method to simulate Raman spectra of $MoS_2$ at realistic defect distributions, yet still retaining DFT accuracy. Sulphur vacancies with concentrations up to 2-3% lead to good agreement with experiments. To decouple effects of oxygen adsorbates, carbonaceous contaminants, and sulphur vacancies, control samples of hBN covered and encapsulated $MoS_2$ are measured. Similar Raman peak evolution with electron dose for hBN covered and uncovered ML $MoS_2$ suggests lack of interaction of carbonaceous contaminants with the sample. Quenching in PL intensity with electron dose at both room and cryogenic temperatures, and emergence of defect-bound excitons at cryogenic temperatures confirms defect formation. We observe narrowing of defect-bound exciton peak linewidth with hBN covering and encapsulation, confirming creation of localized defects ideal for quantum applications. Our work demonstrates a simple approach to create localized defects in ML $MoS_2$ via electron irradiation at ultralow accelerating voltages.

**Results:**

$MoS_2$ flakes were mechanically exfoliated from bulk crystal onto $SiO_2$/Si substrate using scotch tape method. The inset of Fig. 1(b) shows optical image of $MoS_2$ flake on substrate. Layer numbers were determined by measuring flake optical contrast compared to substrate[36]. For pristine ML $MoS_2$, E' and $A_1$' Raman peaks are 386 $cm^{-1}$ and 404 $cm^{-1}$ respectively.

To study defect formation at low accelerating voltages, we irradiated different regions of ML $MoS_2$. We fix accelerating voltage to 3 kV, unless mentioned otherwise. Fig. 1(b) shows scanning electron microscope (SEM) image of an irradiated sample (S1). The yellow box corresponds to an electron dose of $2.6 \times 10^4$ $e^-$ $nm^{-2}$. For details of the dose calculation, see Supporting Information (SI) – III. Carbonaceous contaminants deposited during irradiation result in different contrast for irradiated regions, compared to non-irradiated regions (also confirmed by Raman spectroscopy and atomic force microscopy, for details see SI - VII).

Representative Raman spectra of pristine and electron irradiated samples are presented in Fig. 1(c) and 1(d), respectively. Raman spectrum of the irradiated region (yellow box in Fig. 1(b)) shows redshifts of E' and $A_1$' peaks with respect to the pristine sample even at 3 kV accelerating voltage, well below knock-on voltage for formation of defects in $MoS_2$[14]. Evolution of LO and ZO Raman modes in the irradiated sample (Fig. 1(d)) is attributed to defects. We note that two and three peak fitting were explored initially, but satisfactory fitting was not achieved. It is important to consider LO and ZO peaks as well (see SI - IV). Defect-induced LA-band (~227 cm$^{-1}$) is also observed, but with very low intensity (see SI - VIII), indicating that density of defects created in sample is low.

Fig. 1(e) and 1(f) show PL spectrum of pristine and irradiated ML MoS$_2$, respectively. The overall quenching in PL intensity after irradiation supports the formation of defects[29] and activation of non-radiative decay pathways. X$^0$ intensity is also reduced due to conversion to trions via n-type doping by e-beam. An additional broad peak around 1.6 eV was observed in the irradiated sample (Fig. 1(f)), tentatively labelled as L peak (discussed extensively later). Raman and PL data are also plotted in log-scale to confirm all peaks are accounted for, above the background (see SI - XIX).

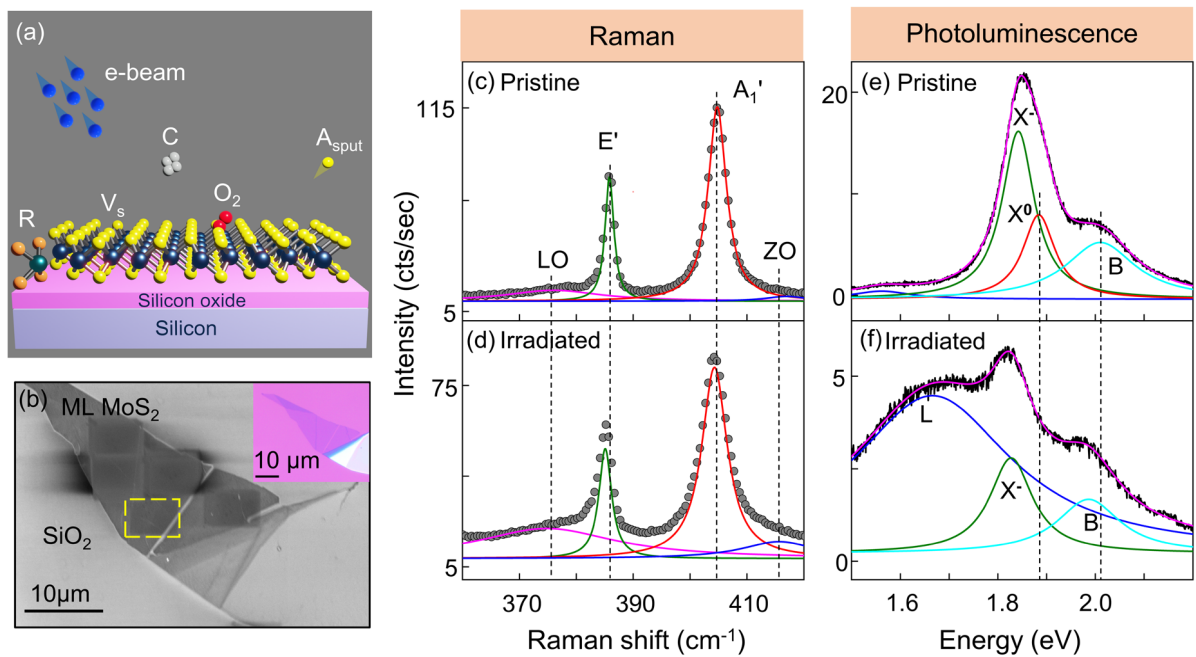

**Figure 1: Raman and photoluminescence (PL) spectra of pristine and electron irradiated ML MoS$_2$.** **(a)** Schematics of electron-beam-induced processes in ML TMDs, including sputtering (A$_{sput}$), radiolysis (R), and carbonaceous contamination (C). Sulphur vacancies and adsorbed oxygen are denoted as V$_s$ and O$_2$, respectively. **(b)** SEM image of the irradiated sample (sample S1) showing e-beam irradiated regions in SEM at 3 kV accelerating voltage.

The yellow box corresponds to an electron dose of $2.6 \times 10^4$ e$^-$ nm$^{-2}$. The inset shows optical micrograph of the pristine MoS$_2$ sample. **(c, d)** Raman spectra of the pristine (c) and irradiated S1 sample ((d), yellow box in (b)). The data is fitted with four Lorentzian peaks (solid lines). **(e, f)** PL spectra of pristine (e) and irradiated S1 samples (f), respectively. The data is fitted with four Lorentzian peaks (solid lines). X$^0$, X$^-$, B, and L denote neutral A exciton, negative A trion, B exciton, and low energy broad peak, respectively. All spectra were measured at room temperature.

To understand defect formation mechanisms, we recorded a series of Raman spectra (Fig. 2(a)) of ML MoS$_2$ irradiated at electron doses ranging from 0$\xi$ to 18$\xi$ ($\xi = 1.3 \times 10^4$ e$^-$ nm$^2$, 0 = pristine sample). Raw Raman spectra for all doses are presented in SI - V. The fitting of E', A$_1$', LO and ZO peaks are indicated in Figure 2a. The inset shows integrated LA band intensity normalized with integrated E' peak intensity, showing minor contribution of LA band, which increases with dose (see SI VIII for details). The evolution of $k$, $\Gamma$, wavenumber separation ($\Delta k$), and intensity ratio of E' and A$_1$' peaks with electron dose are tracked. To better understand the evolution of these parameters, we focus on two different dose ranges, (I) 0 to 2$\xi$, and (II) 2$\xi$ to 18$\xi$. In dose range I, we observed redshift of $k$, and increased $\Gamma$ of both E' and A$_1$' peaks (Fig. 2(b), 2(c)). To achieve even lower dose, irradiation was carried out at different magnifications while keeping the same beam current (see SI XI). The small (≤ 0.2 cm$^{-1}$) variations in FWHM of first-order Raman peaks of pristine samples (Figure 2c) can be attributed to the sample preparation process and local substrate variations. For example, we observed ~ 0.1 cm$^{-1}$ spatial variations in FWHM across a single sample attributed to variations of local environment. For a particular pristine sample, the indicated FWHM is the mean of FWHMs of different points in the sample. Please also see Raman measurements on other samples in SI – XXI. Further, to study spatial variation in electron irradiated areas, spatial mapping of frequency and FWHM of E' and A₁' Raman modes were also performed (see SI – XXII for details).

The changes in E' and A₁' peaks are attributed to defect formation[15,32] and electron doping[27] respectively. For dose range II, E' peak continues to redshift, attributed to increased sulphur vacancies. In contrast, A₁' peak blue shifts with decreased $\Gamma$, which is attributed to reduced n-type doping due to removal of oxygen atoms from sulphur vacancy sites, and resulting increase of p-type doping due to increased sulphur vacancies. The role of carbonaceous contaminations in Raman spectra evolution is minor, and is discussed later. The increase in peak separation ($k_{A_1'} - k_{E'}$) with electron dose (SI - VI) can be attributed to an overall rise in sulphur vacancies[15]. However, if defect formation and doping effects are simultaneously present, separation of E' and A₁' peaks may not be a good measure of defect formation (also see SI – VI).

To understand accelerating voltage-dependent defect creation, we used two different accelerating voltages (3 and 5 kV). We did not observe any qualitative differences in 3 kV (Sample S1) and 5 kV (Sample S2, SI - II) irradiated samples, suggesting similar defect formation mechanisms (also see SI – VI). The intensity ratio ($I_{E'}/I_{A_1'}$) of E' and A₁' modes reduce more for 5 kV irradiation, as compared to 3 kV irradiation (SI - VI). $\Gamma$ increase is also higher for 5 kV irradiation, compared to 3 kV irradiation. In contrast, wavenumber separation ($\Delta k$) change with electron dose for 3 kV and 5 kV irradiated samples are similar (SI - VI). Overall, irradiation at 5 kV seems to cause slightly more damage than 3 kV irradiation, as also evidenced by room temperature PL data (SI - XX). We note that the comparison of 3 kV vs 5 kV is influenced not just by displacement cross-sections, but by beam width and beam penetration. Suspending samples on a TEM grid may help alleviate this issue, but are beyond the scope of this study.

Recent studies have discussed the reduction of displacement threshold energy of sputtering atoms due to localization of electronic excitations to emerging defect sites[19]. Adsorbed oxygen atoms on sulphur vacancy sites may result in lowering the threshold for damage, by increasing localization of electronic excitations. Further, for ionization and radiolysis processes, the energy transferred to the medium increases as accelerating voltage decreases, scaling as $1/E$, then levelling off and eventually decreasing at low acceleration voltages of about 1 kV[37]. Since ultralow accelerating voltages are interesting for a growing number of researchers, further theoretical studies are sorely needed for understanding the competing defect formation mechanisms at these voltages (also see SI - XVIII for more discussion on other mechanisms).

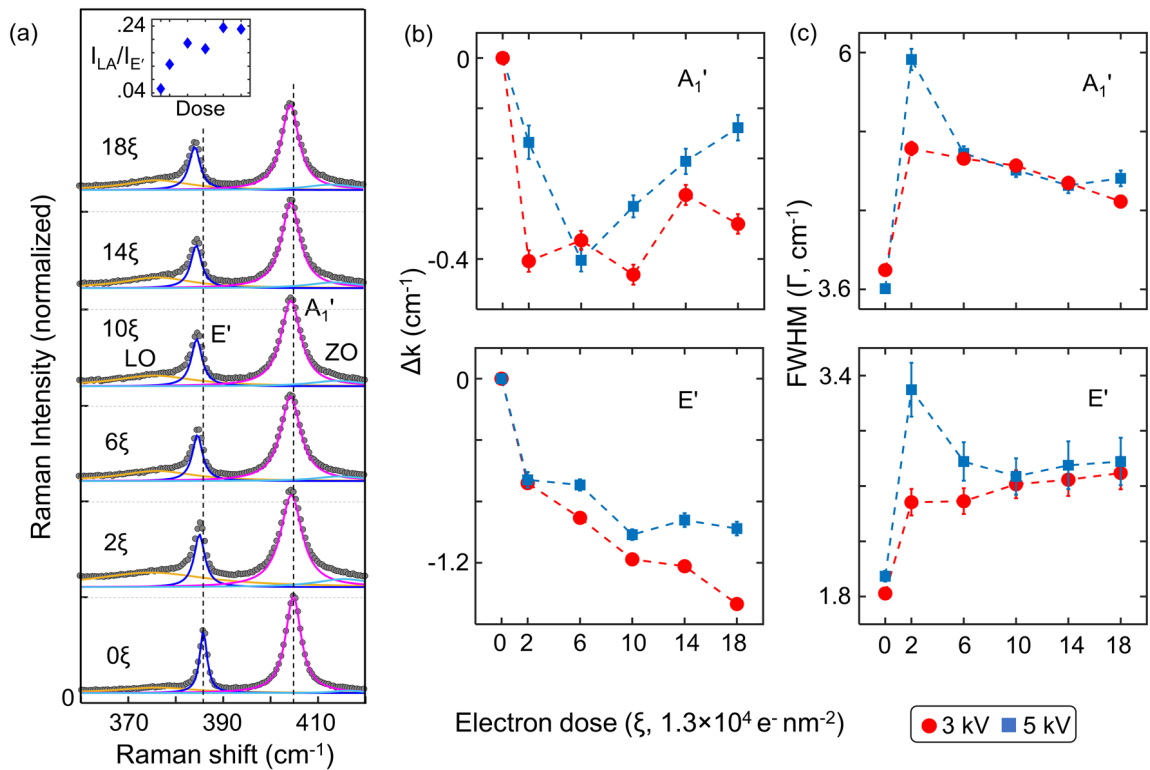

**Figure 2: Evolution of Raman spectra of ML MoS$_2$ with electron dose.** Series of Raman spectra of the pristine and irradiated sample irradiated at 3 kV (S1) and 5 kV (S2) accelerating voltage. **(a)** 3 kV data shown with varying doses, along with fitting of E', A$_1$', LO and ZO peaks. All spectra are normalized with respect to the Si Raman peak. Variation of Raman peak

parameters with electron dose **(b)** peak shift from pristine sample ($\Delta k$), **(c)** full width at half maximum (FWHM, $\Gamma$) of E' and $A_1$' modes. These parameters are obtained by fitting four Lorentzian peaks.

To investigate direction and magnitude of the shifts upon introduction of defects, we carried out simulations of Raman spectra based on DFT. Modelling effects of randomly distributed defects on vibrational spectra requires large supercells and to this end, we have adopted two advanced methods. First, MLFF were trained using small defective systems and with those, the vibrational modes of large supercells could be readily evaluated (for details see SI – I and SI - XIV). Second, Raman tensors of the supercell modes were obtained using RGDOS method, which relies on the projection to unit cell modes and summing up the corresponding Raman tensors. The simulated spectra for three defects, S vacancy, O substitutional in S-site, and O adatom on top of S, at 5% defect concentration are shown in Fig. 3(a). The simulated spectra for other concentrations, along with added discussion are given in SI – XV and SI - XVI. In addition, the evolution of the peak positions as a function of % of unit cells with defects are shown in Fig. 3(b).

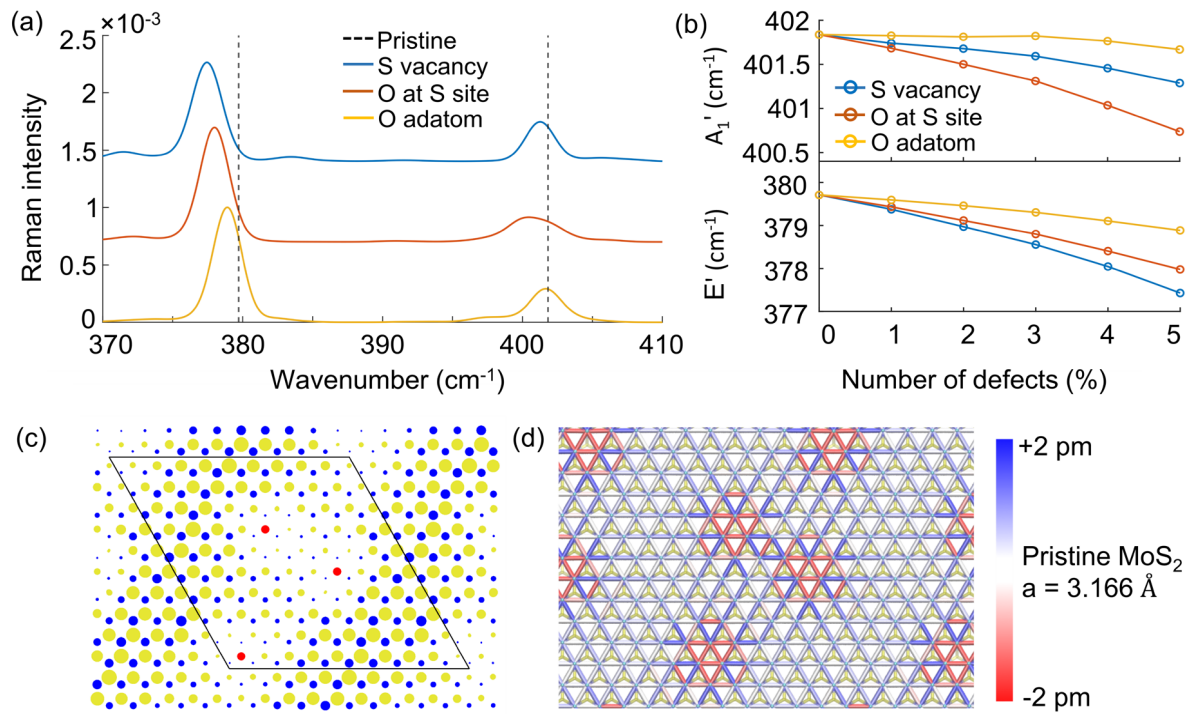

**Figure 3: Simulated Raman spectra. (a)** Simulated spectra of $MoS_2$ at 5% defect concentration (S vacancy, O at S site, or O adatom), and comparison to the peak positions in pristine $MoS_2$. **(b)** Raman peak shifts as a function of % of unit cells with defects. **(c)** Illustration of the spatial mapping of E' vibrational mode amplitude in the case of 3% S vacancies. The size of the circle denotes the magnitude of the eigenvalue in Mo (blue) and S (yellow) atoms. Position of the S vacancies are highlighted with red circles. **(d)** Strain distribution in the case of 3% S vacancies. Mo-Mo bonds are coloured by the bond length (or lattice constant) relative to the pristine $MoS_2$ value. The atomic structure is also shown underneath.

For S vacancy, E' peak downshifts fairly strongly, which agrees with experimental observations. Quantitatively, 1-1.4 cm$^{-1}$ downshift of E' peak would correspond to about 3-4 % vacancy concentration, which seems reasonable under irradiation conditions considered herein. The E'-mode eigenvector and strain distribution for the case of 3% S vacancies are shown in Fig. 3(c, d). The eigenvector is seen to be localized to regions away from the vacancy.

Therein the lattice is tensile-strained, owing to lattice contraction at the vacancy, and this tensile strain translates to a decrease in E' frequency (SI - XVI). The $A_1'$ peak downshifts much less when defects are introduced, which agrees with the much smaller shift with strain. Part of the initial downshift seen in experiments could also arise from vacancies, but the eventual change in direction of the shift cannot, and thus we think this is explained by the doping as discussed above. Extracted peak widths (SI - XVI) indicate stronger broadening for E' peak than $A_1'$ peak, which also qualitatively agrees with experiments.

We also consider a scenario where the sample initially contains O at S-site, which irradiation could effectively turn into S vacancies. Since the downshift of E' peak for O at S-site is only slightly smaller than for S vacancy (0.5 cm$^{-1}$ less for 5% defect concentration), such conversion cannot solely explain the experiments, as that would necessitate unrealistically high O concentration of more than 10%. Moreover, that would be accompanied by a large upshift of $A_1'$ peak, inconsistent with experiments. Alternatively, part of the redshift of E' peak could also arise from O adatoms remaining on the surface after leaving the S-site. The peak shift from adatoms is relatively small but should contribute additively when both adatoms and S vacancies are present.

To decouple effects of carbonaceous contaminants and substrate from doping and defect creation, we prepared hBN covered and encapsulated samples (see SI - I and II for methods and optical images). Thickness of hBN (∼ 8 nm) is chosen to allow e-beam interaction with MoS$_2$, while restricting carbonaceous contaminants deposition on MoS$_2$.

Fig. 4 compares E' and $A_1'$ peak changes with electron dose for bare, hBN covered and hBN encapsulated ML MoS$_2$. The possible interactions of these different samples with contaminants and substrate are schematically presented in Fig. 4(a-c). $k$ and $\Gamma$ for E' and $A_1'$ Raman peaks in dose range I and II (Fig. 4(d-g)) have similar trends for bare and covered MoS$_2$, suggesting lack of interaction of carbonaceous contaminants with $MoS_2$. Hence, decrease in $\Gamma$ of $A_1'$ peak in dose range II is not due to carbonaceous contaminant doping effects. Instead, carbonaceous contamination results in increasing Raman background (SI - VII).

Sulphur vacancies in ML $MoS_2$ are shown to be acceptors[19] and can cause p-type doping[38–40]. In contrast, oxygen adsorption can fill the S-vacancy site, and result in the usually observed n-doping in $MoS_2$. Then, removal of oxygen atoms by e-beam can cause reduced n-type doping. As a result, at higher irradiation time, p-type doping due to sulphur vacancies and removal of oxygen atoms dominates over n-type doping due to e-beam, and causes blue shift and decreased $\Gamma$ of $A_1'$. We note that evolution of E' and $A_1'$ peak shifts for hBN encapsulated $MoS_2$ is not consistent across various samples, whereas $\Gamma$ is consistent (SI – IX). The various doping processes involved in the electron irradiation of $MoS_2$ and their contribution for dose range I and II are illustrated in SI - XII. Further, Monte Carlo simulations of electron trajectories are performed at 3 kV accelerating voltage using CASINO software[41], providing insights into interaction of e-beam with ML $MoS_2$ in presence of hBN and carbonaceous contamination (SI - X), and backscattered electrons (SI – XVII).

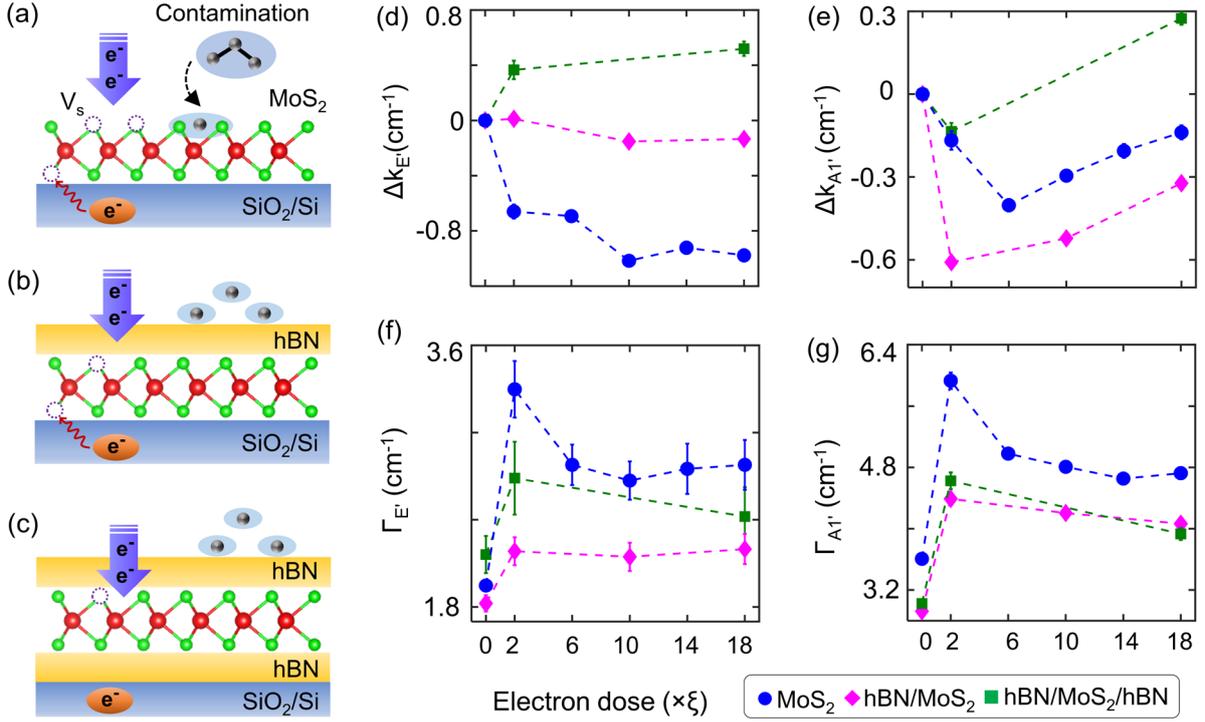

**Figure 4: Evolution of Raman spectra of bare, hBN covered, and encapsulated ML MoS$_2$ with electron dose. (a-c)** Schematics showing the effect of hBN protection and carbonaceous contamination deposition on defect formation and doping in irradiated ML MoS$_2$. **(d), (e)** show variation in peak positions ($\Delta k$), full width at half maximum (FWHM, $\Gamma$) of E' and A$_1$' Raman peaks with electron dose, respectively. $\Delta k$ again indicates that the wavenumbers of E' and A$_1$' peaks are subtracted from their respective wavenumbers in the pristine sample. The change in full width at half maximum (FWHM, $\Gamma$) of E' and A$_1$' Raman peaks with electron dose are shown in **(f)** and **(g)**, respectively. These parameters were obtained by fitting four Lorentzian peaks to the data. Electron irradiation is performed at 5 kV accelerating voltage. Here, $\xi = 1.3 \times 10^4$ e$^-$ nm$^2$.

PL spectroscopy gives direct insights into the MoS$_2$ excitonic states, including intra-gap defect-bound excitons. To decouple defect formation and doping effects, we measured cryogenic PL (4K) of the three irradiated samples i.e., bare, hBN covered, and encapsulated ML MoS$_2$. The cryogenic PL of pristine bare ML MoS$_2$ (Fig. 5(a), 0$\xi$) shows X$^0$ and X$^-$ peaks, and a broad PL peak (L, 1.75 eV) corresponding to adsorbates bound to chalcogen vacancies[42]. A slight exposure of e-beam or mild annealing can remove the adsorbates and open the $MoS_2$ surface for defect formation. Evolution of PL spectra with electron dose for irradiated bare ML $MoS_2$ is shown in Fig. 5(a). The quenching of PL intensity with electron dose is attributed to defect formation and activation of nonradiative decay channels.

The cryogenic PL spectrum of pristine hBN covered $MoS_2$ (Fig. 5b, $0\xi$) shows well distinguished $X^0$ and $X^-$ peaks, along with a diminished L peak, indicative of high sample quality. For low electron dose ($\delta\xi$), reduction in $X^0$ and $X^-$ peak intensity is observed, along with increase in broad L peak. With further increase in electron dose, high PL quenching is attributed to increase in defect density. The PL spectrum of pristine encapsulated ML $MoS_2$ (Fig. 5(c), $0\xi$) shows $X^0$ peak along with low intensity broad L peak. The absence of $X^-$ peak is due to blockage of charge transfer from $SiO_2$/Si substrate by bottom hBN layer. For low electron dose (Fig. 5(c), $\delta\xi$), the $X^0$ PL intensity is quenched along with appearance of $X^-$ peak, attributed to defect formation and electron doping respectively. With further increase in electron dose, the excitonic PL continues to quench and relative intensity of L peak increases due to increase in sample defect density. However, the extent of damage (compared to bare and hBN covered ML $MoS_2$) has reduced due to hBN encapsulation[43]. Further, inset of Fig. 5(c) shows sharp defect peaks in irradiated encapsulated samples. The hBN layer protects the created defect sites from surrounding environment and substrate, and thus, defect density and linewidth of defect-bound excitons are reduced. These sharp peaks may be promising for SPEs, but detailed studies are beyond the scope of this paper. Power and temperature dependent studies signifies bound nature of the L peak (SI - XIII). L peak shows linear dependence with laser power for bare and hBN covered samples (due to high density of defects), but saturates for hBN encapsulated samples. Also, the intensity of L peak reduces with temperature, signifying the defect origin. Further, appearance of L peak in PL spectra of all irradiated samples (bare, hBN covered, and encapsulated ML MoS$_2$) confirms that L peak corresponds to excitons bound to sulphur vacancies (or other defects) and the interaction of carbonaceous contaminants with sample is negligible.

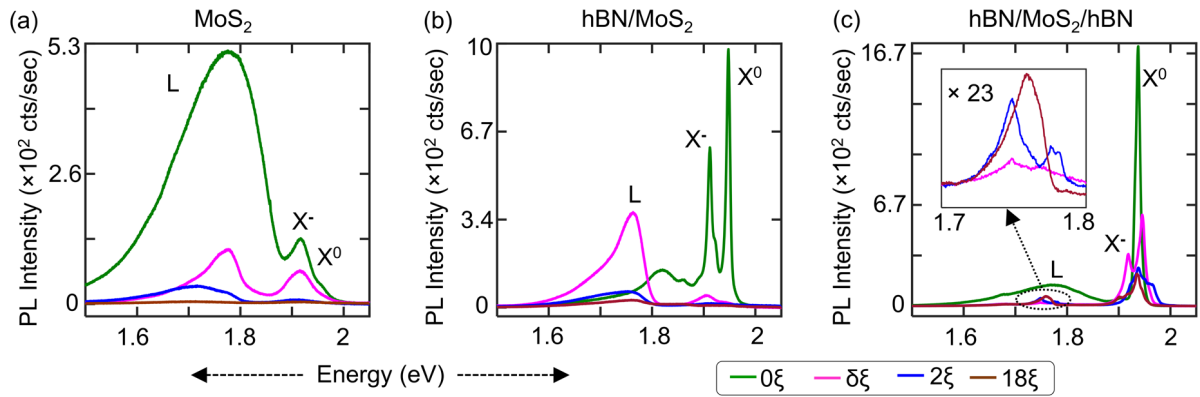

**Figure 5: Cryogenic photoluminescence (PL) of irradiated bare, hBN covered, and hBN encapsulated ML MoS$_2$.** Evolution of cryogenic PL with electron dose for bare, hBN covered, and encapsulated ML MoS$_2$ are presented in **(a)**, **(b)**, and **(c)**, respectively. All spectra are taken at 4K temperature. Electron irradiation is performed at 5 kV accelerating voltage. Some regions of the sample are not intentionally irradiated, however receive an average electron dose of $0.1\xi$ during SEM (denoted as $\delta\xi$). Inset of (c) denotes the zoomed in spectra of L peak, indicating sharp defect peaks. Here, $\xi = 1.3\times10^4$ e$^-$ nm$^2$.

**Conclusion:**

We have demonstrated defect formation in ML MoS$_2$ by e-beam irradiation at ultralow electron accelerating voltages (3-5 kV), and characterized the defects using Raman and PL spectroscopy. The evolution of E' and A$_1$' Raman modes and overall quenching in PL intensity indicate defect formation. Defect formation also activates ZO and LO Raman peaks. We have developed a model for simulating Raman spectra in low defect density samples, using a combination of DFT and MLFF calculations, with good agreement with experiments. Further, we decoupled doping effects due to carbonaceous contaminants, defect creation and e-beam using hBN covered and encapsulated $MoS_2$ control samples. Carbonaceous contaminants do not participate in defect creation or contribute to PL. Interestingly, defect density and corresponding PL linewidth of defect-bound excitons can be reduced via hBN covering and encapsulation. Our work demonstrates a novel approach for creating isolated defects using ultralow electron accelerating voltages in 2D materials ideal for quantum applications. We hope our work will motivate researchers to perform further studies for understanding the defect formation mechanism at ultralow electron accelerating voltages[44].

ASSOCIATED CONTENT

The following files are available free of charge.

Supporting Information (PDF)

AUTHOR INFORMATION:

**Corresponding Author:**

*Akshay Singh, aksy@iisc.ac.in

**Author Contributions**

AKD and AS developed the experimental framework. EB, TL and HPK performed the DFT and MLFF modelling of defects. AKD performed the optical experiments and electron beam irradiation, with assistance from HS, MM and PRP. AKD performed the data analysis, with assistance in PL analysis by HS. KW and TT provided the hBN bulk crystals. AKD, HPK and AS discussed and prepared the manuscript, with contributions from all authors.

**Data Availability**

The data that support the findings of this study are available upon request from the authors.

**ACKNOWLEDGMENTS**

AS would like to acknowledge funding from Indian Institute of Science start-up and SERB grant (SRG-2020-000133). AKD would like to acknowledge Prime Minister's Research Fellowship (PMRF). The authors also acknowledge Micro Nano Characterization Facility (MNCF), Centre for Nano Science and Engineering (CeNSE) for use of characterization facilities. The authors also acknowledge Arindam Ghosh Lab (IISc) for use of Raman facility. TL acknowledges financial support from Infotech Oulu Doctoral Program. We gratefully acknowledge CSC–IT Center for Science, Finland, for computational resources. T.T. acknowledges support from the JSPS KAKENHI (Grant Numbers 19H05790 and 20H00354) and A3 Foresight by JSPS.

**ABBREVIATIONS**

2D, two dimensional; TMDs, transition metal dichalcogenides; ML, monolayer; SPEs, single-photon emitters; SEM, scanning electron microscope; FIB, focused ion beam; TEM, transmission electron microscope; e-beam, electron beam; PL, photoluminescence; DFT, density functional theory; MLFF, machine learning force fields.

Supporting Information

# Evidence of defect formation in monolayer MoS$_2$ at ultralow accelerating voltage electron irradiation

Ajit Kumar Dash[1], Hariharan Swaminathan[1], Ethan Berger[2], Mainak Mondal[1], Touko Lehenkari[2], Pushp Raj Prasad[1], Kenji Watanabe[3], Takashi Taniguchi[4], Hannu-Pekka Komsa[2], Akshay Singh[1, *]

[1]Department of Physics, Indian Institute of Science, Bengaluru, Karnataka -560012, India

[2]Microelectronics Research Unit, University of Oulu, FI-90014, Oulu, Finland

[3]Research Center for Functional Materials, National Institute for Materials Science, Ibaraki 305-0044, Japan

[4]International Center for Materials Nanoarchitectonics, National Institute for Materials Science, Ibaraki 305-0044, Japan

*Corresponding author: aksy@iisc.ac.in

**(I) Methods:**

**Sample preparation:** MoS$_2$ (purchased from 2D semiconductors) and hBN flakes were prepared by micro-mechanical exfoliation of respective bulk crystals using scotch tape method. We used the PDMS-PPC-based transfer method to prepare hBN covered and hBN encapsulated ML MoS$_2$ samples. A PDMS dot (~300 $\mu m$) was prepared on a glass slide by mixing Sylgard 184 PDMS base and curing in the 10:1 ratio. Then, polypropylene carbonate (PPC) solution (15% solution of PPC in Anisole) was spin-coated on cured PDMS dot. The PDMS-PPC dot was used to pick up and drop the hBN on 1L MoS$_2$ in a customized transfer setup. Maximum temperature used during transfer was 110°C to avoid oxidation of the sample. After transfer, the hBN/MoS$_2$ stack was treated with anisole (2min), acetone (2min) and IPA (2min) to remove PPC residue. Then, the sample was annealed in an inert atmosphere for 3 hours at 250 °C to remove organic contaminants and enhance the heterostructure interface. Same procedure was followed to prepare hBN encapsulated ML MoS$_2$ sample.

**Electron beam irradiation:** Electron beam irradiation and secondary electron (SE) imaging were performed in a Zeiss Ultra 55 scanning electron microscope. The electron dose was varied by scanning regions for different times at fixed accelerating voltage and specimen current (for details, see SI - III). Two accelerating voltages of 3 kV and 5 kV were used for irradiation. Imaging of irradiated samples was done in InLens mode at lower magnification to reduce dose per unit area.

**Raman and PL measurements:** Raman measurements were carried out in a HORIBA LabRamHR Raman setup using a 532nm laser, 100x objective, and 1800 lines/mm grating. Laser power was kept low ($\sim 100 \mu W$) to avoid laser-induced damage. Room temperature PL measurements were performed in a Witec Alpha 300 system using a 532nm laser, 100x objective, and 600 lines/mm grating. Cryogenic PL measurements were performed at 4K using customized set up made up of Montana cryosystem, Andor spectrometer (300 grating lines/mm) and silicon CCD. All low temperature (4K) measurements were recorded using a 482 nm excitation laser, 50X objective and power $\leq 100 \mu$W.

**DFT and MLFF calculations:** All DFT calculations, training of the MLFF, and the calculation of vibrational modes using MLFF molecular dynamics simulations were carried out using Vienna ab initio simulation package (VASP)[45,46]. In particular, the MLFF was generated using on-the-fly learning scheme during DFT MD run, as described in Refs.[34,35]. The final MLFF was built by running separate MD simulations of defective and strained systems and compiling their MLFFs into single supermodel. In total, the supermodel contains information on the three relevant defect configurations and ±2% strain (see Supporting Information - XIV, XV, XVI for details). In all DFT calculations, we are using rev-vdW-DF2 van der Waals functional[47] and a plane wave basis of 500 eV to represent the electronic wavefunctions. In MD calculations, a canonical ensemble was simulated by a Nose-Hoover thermostat[48,49]. The Raman tensors of the MoS$_2$ primitive cell, as required for the RGDOS calculation, were evaluated from the change of dielectric constant with the phonon displacement via finite-differences.

**(II) Optical and SEM images of bare and hBN covered ML MoS$_2$:**

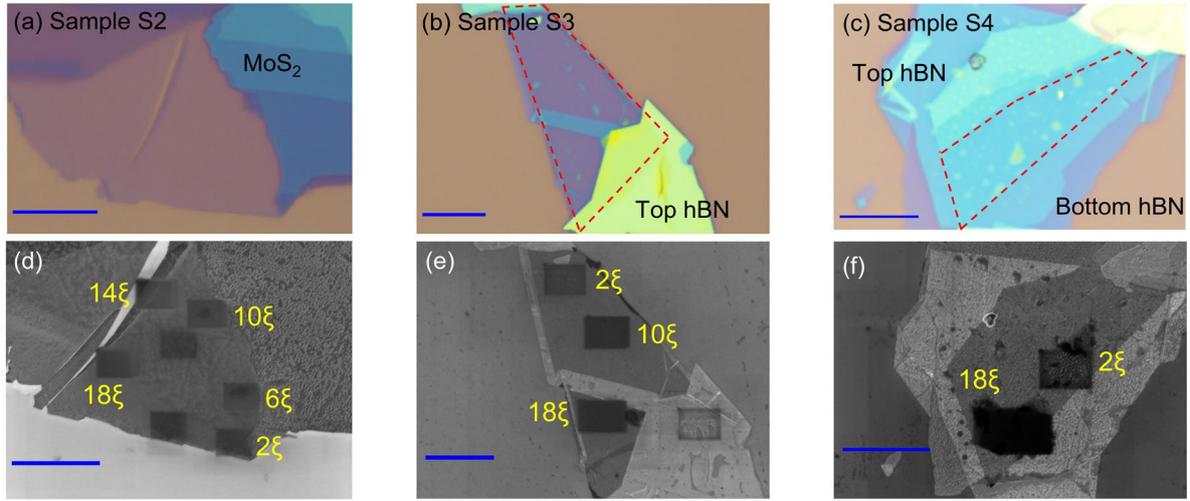

**Figure S1: (a)**, **(b)**, **(c)** are optical images of bare, hBN covered, and hBN encapsulated ML MoS$_2$, respectively. Corresponding SEM images after e-beam irradiation are shown in **(d)**, **(e)**, and **(f)**. All scale bars are 10$\mu$m. Here, $\xi = 1.3\times10^4$ e$^-$ nm$^2$.

**(III) Calculation of electron dose of irradiation in SEM:**

The number of electrons that sample receives per unit area due to irradiation is given by, $e^- \, dose = \frac{I \times t}{e \times A}$, where $I, A$ stand for specimen current and area scanned on the sample, respectively, $t$ stands for total irradiation time, and $e$ is the electronic charge. The area scanned on the sample ($A$) can be calculated using the magnification ($M$) formula, $M = \frac{L_I}{L_S}$, $A_S = (L_S)^2$, where $L_I, L_S$ denotes the length of the imaging screen and the scanned region, respectively. In a scanning electron microscope (SEM), the electron dose can be varied by changing $I, t,$ or $A$. In our experiments, electron dose scales with the irradiation time for fixed $I$ and $A$. Electron dose can also be varied by changing $I$, proportional to the aperture size. However, it is challenging to irradiate the sample's different regions with different doses by changing $I$, as the sample's focus changes by changing the aperture. Another method for varying electron dose in SEM is changing the area scanned on the sample (related to $M$). But it is not preferable for irradiating mechanically exfoliated monolayer (ML) samples because it requires large-size flakes.

**(IV) Comparison of two, three, and four peak fitting of Raman spectra of electron irradiated monolayer MoS$_2$:**

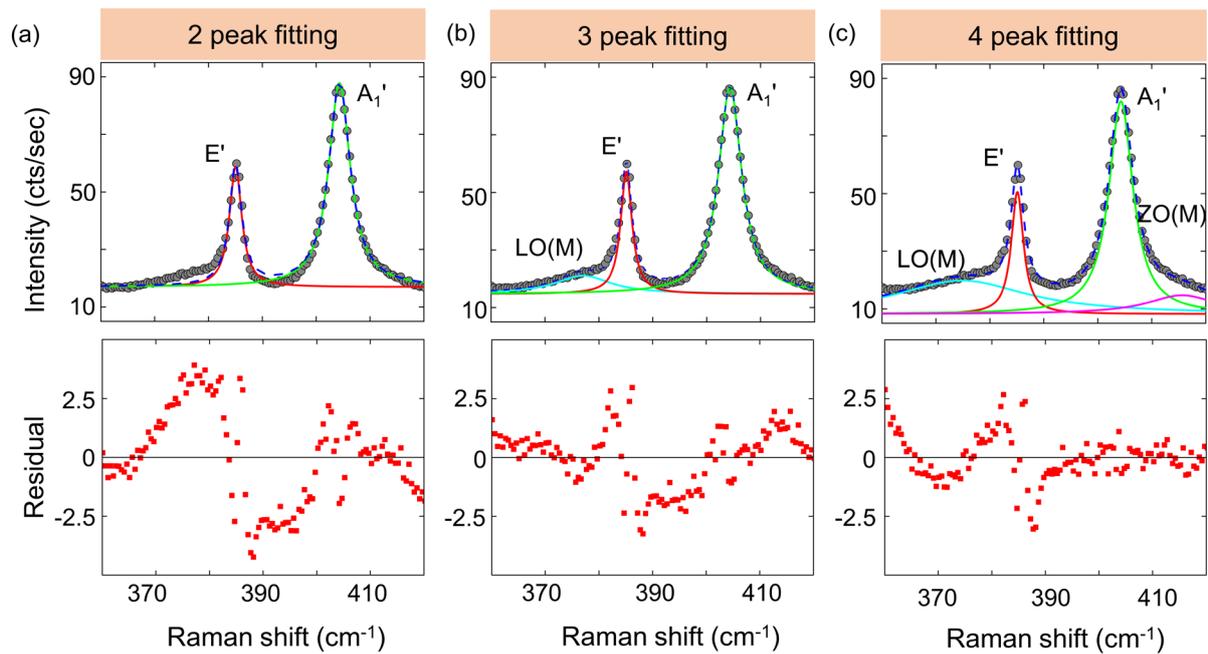

**Figure S2:** Comparison of two, three, and four peak Lorentzian fitting for Raman spectra of electron irradiated monolayer MoS$_2$ shown in Fig. 1(d). Respective residuals of fitting are also plotted below.

We have fitted the Raman spectra of irradiated sample by a four peak Lorentzian function that considers LO + E' + $A_1'$ + ZO peaks. LO and ZO Raman peaks have been observed earlier for ion/electron beam irradiated monolayer $MoS_2$ [Ref. 15, 16, 18]. In our own work, two and three peak fitting were explored initially, but satisfactory fitting was not achieved. Figure S2 compares two, three and four peak Lorentzian fitting of Raman spectra irradiated sample.

**(V) Evolution of Raman spectra of ML $MoS_2$ with electron dose:**

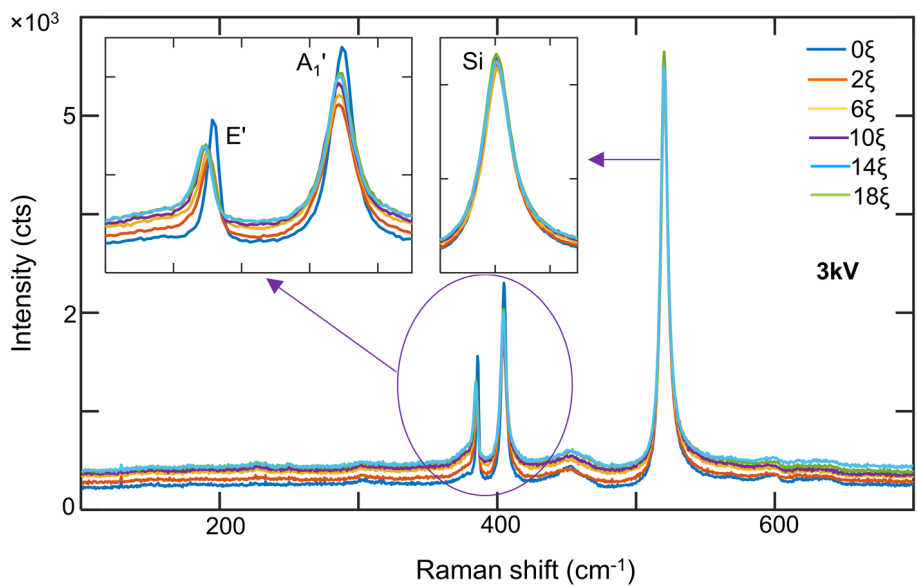

**Figure S3a:** Evolution of Raman spectra of ML $MoS_2$ (Sample, S1) with electron dose ($\xi = 1.3 \times 10^4$ e⁻nm⁻²). Irradiation is performed at 3 kV accelerating voltage.

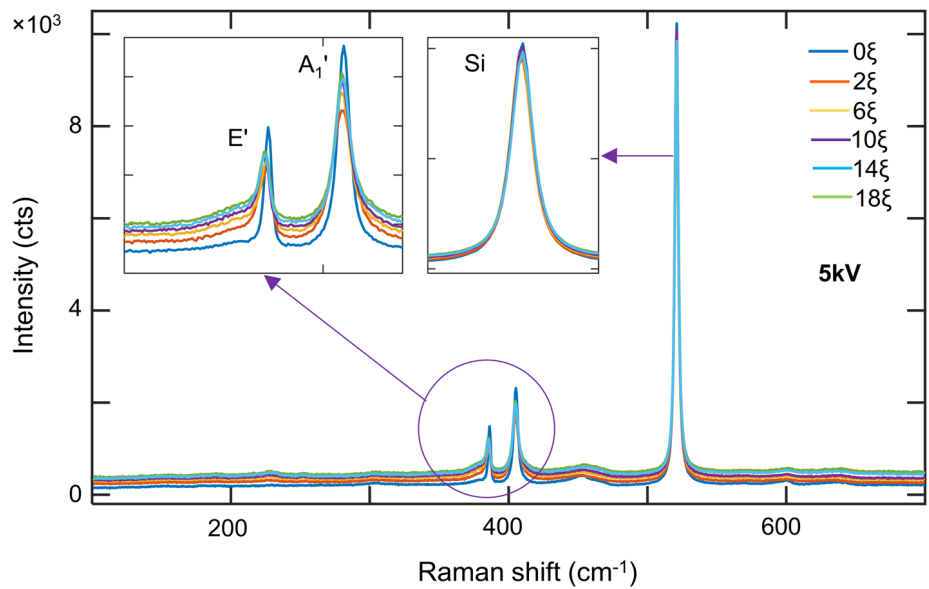

**Figure S3b:** Evolution of Raman spectra of ML MoS$_2$ (Sample, S3) with electron dose ($\xi = 1.3 \times 10^4$ e$^-$nm$^{-2}$). Irradiation is performed at 5 kV accelerating voltage.

## (VI) Evolution of Raman peak parameters of ML MoS$_2$ with electron dose:

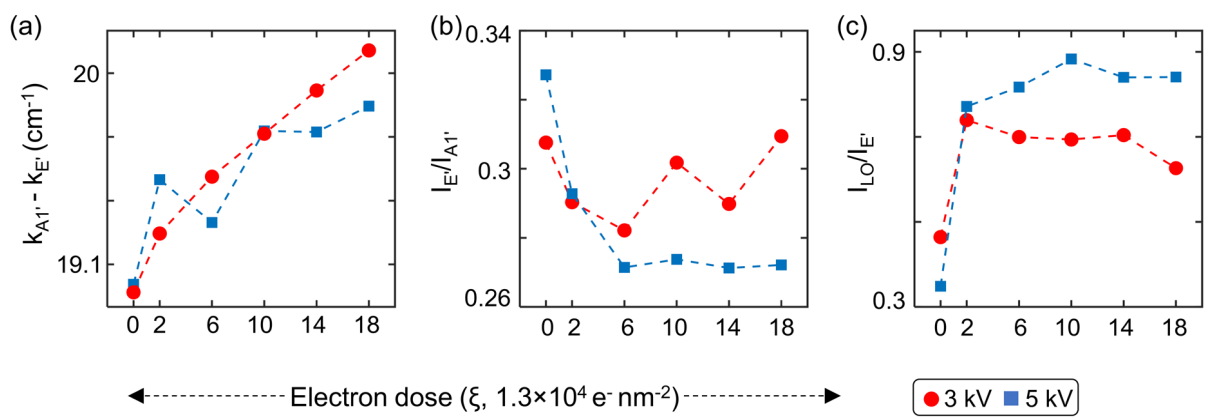

**Figure S4: (a)** Variation in peak separation ($k_{A_1'} - k_{E'}$), **(b)** intensity ratio of E' and A$_1$', and **(c)** intensity ratio of LO and E' modes are plotted against electron dose for ML MoS$_2$, irradiated at 3kV and 5kV electron accelerating voltages.

## (VII) Carbonaceous nature of contaminants deposited during e-beam irradiation:

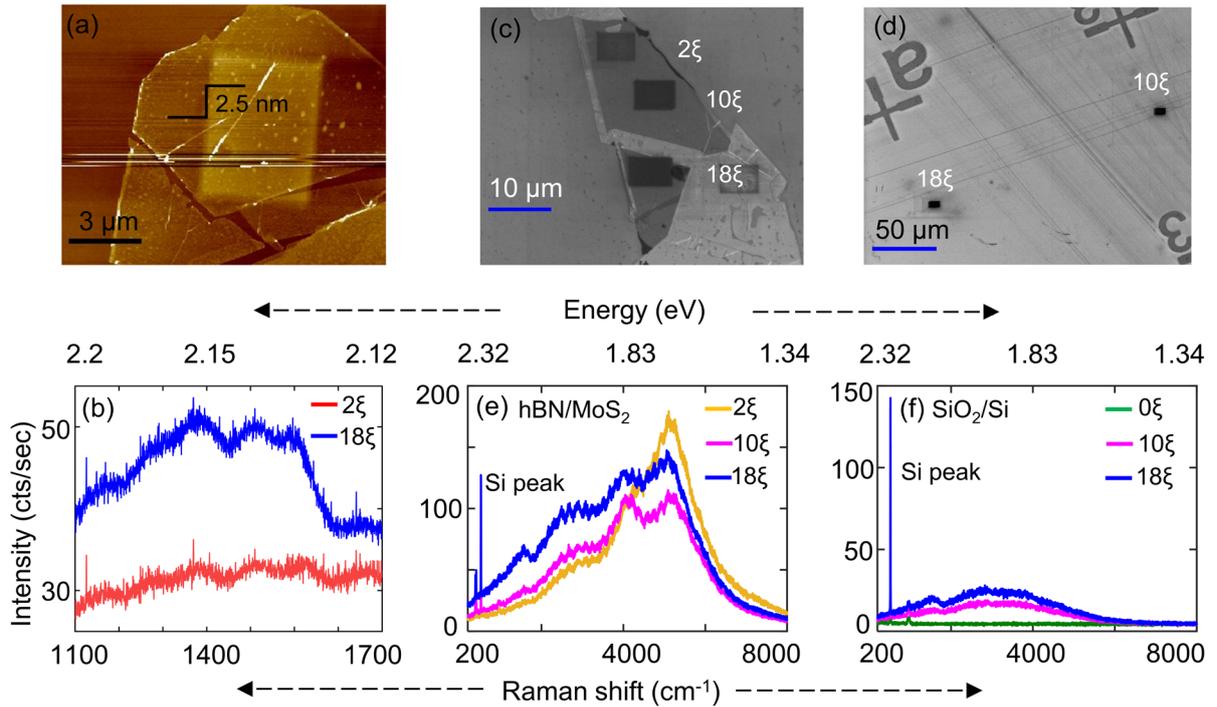

**Figure S5: (a)** Atomic force microscope (AFM) image of electron irradiated ML MoS$_2$ (Sample, S4) clearly shows contaminant depositions in the e-beam exposed region. **(b)** Raman spectrum electron irradiated ML MoS$_2$ from 1100-1800 cm$^{-1}$ shows two amorphous carbon peaks. **(c), (d)** SEM image of electron irradiated hBN/MoS$_2$ and silicon substrate, respectively. **(e), (f)** Larger range Raman spectra of irradiated hBN/MoS$_2$ and silicon substrate respectively. Irradiation is performed at 5 kV accelerating voltage. Here $\xi = 1.3 \times 10^4$ e$^-$ nm$^{-2}$.

In the presence of e-beam, the carbonaceous contaminants present in the SEM chamber break down into free radicals and are deposited onto the sample surface. The carbonaceous contaminants result in different contrast for the exposed regions, compared to unexposed regions. The carbonaceous contaminants have a Raman response in the range 1000 cm$^{-1}$ to 1800 cm$^{-1}$ (Fig. S5a), demonstrating two peaks corresponding to amorphous carbon. We further observed an increase in the background of the Raman spectrum with the electron dose for MoS$_2$, hBN covered MoS$_2$ sample and silicon substrate (Fig. S5e, S5f). In our experiments, dose scales with irradiation time (SI-III), and the amount of deposited contamination increases with time. For irradiated silicon substrate, a broad PL peak is observed and the intensity of the broad peak increases with electron dose (S5f). The tail of the broad PL peak extends to Raman range of $MoS_2$, and cause overall upshift of the Raman spectrum. Thus, increase in the Raman spectrum background with increasing electron dose is due to carbonaceous deposits. Fig. S5e indicates the PL contribution (defect, excitons) to Raman response.

**(VIII) Defect related LA-band in electron irradiated ML $MoS_2$:**

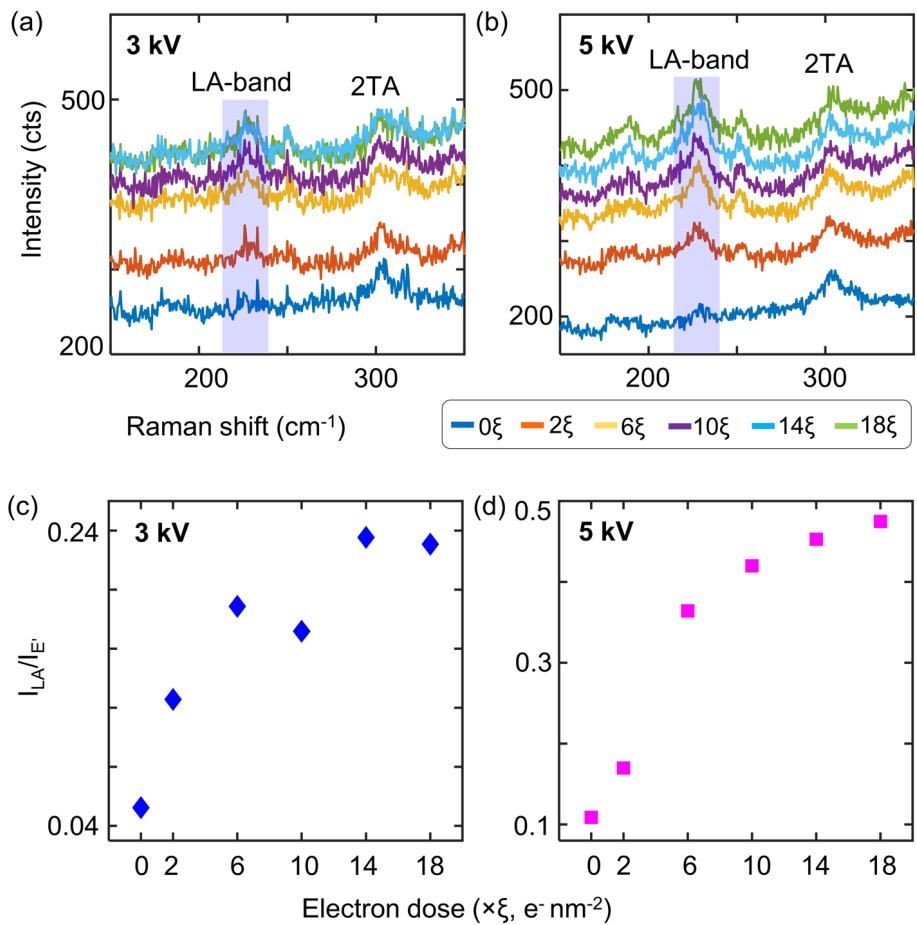

**Figure S6:** Evolution of LA-band of monolayer $MoS_2$ with electron dose for **(a)** 3kV and **(b)** 5kV electron irradiation. The ratio of integrated intensity of LA-band and E' Raman mode with electron dose for 3kV and 5 kV electron irradiation are plotted in **(c)** and **(d)**, respectively. The wavenumbers over which LA band is integrated is indicated by the shaded boxes in (a, b). Here $\xi = 1.3 \times 10^4$ $e^-$ $nm^{-2}$.

As we are using ultralow accelerating voltage electron irradiation, the density of defects created in the sample is low. As shown in Fig. S6, there is minor contribution from LA band to the Raman spectrum. On the other hand, in Raman spectrum of irradiated samples, the LA-band seems to be increasing with irradiation time, which is consistent with increasing density of defects.

**(IX) Raman peak properties electron irradiated encapsulated ML MoS$_2$ samples:**

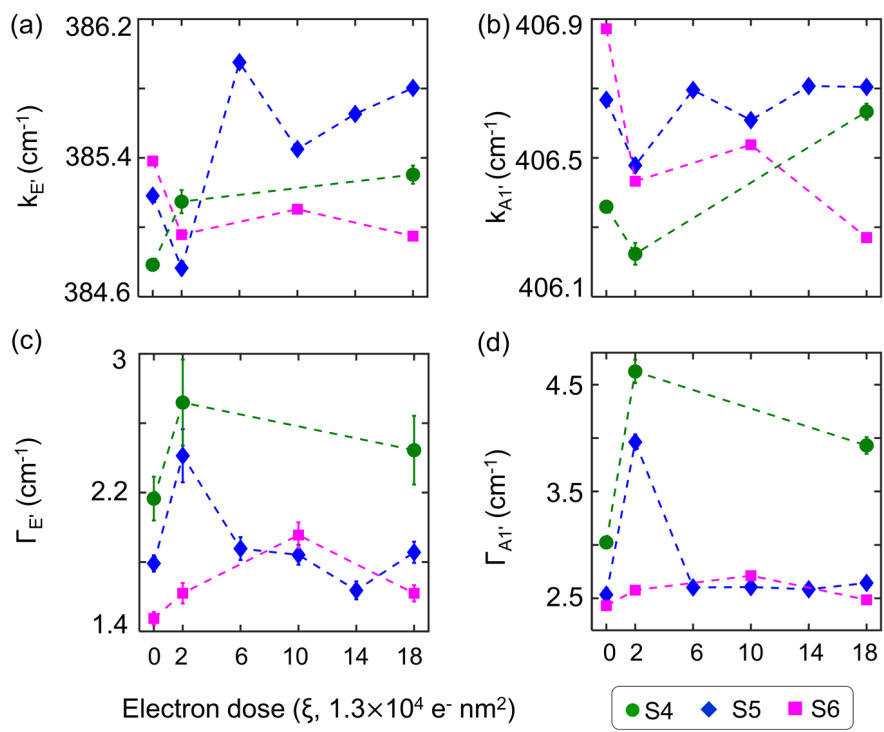

**Figure S7: (a)**, **(b)** Evolution of E' and A$_1$' peak positions (k) of encapsulated ML MoS$_2$ samples (S4, S5, S6) with electron dose. **(c)**, **(d)** Variation of E' and A$_1$' full width half at maximum ($\Gamma$) of encapsulated ML MoS$_2$ samples (S4, S5, S6) with electron dose. Electron irradiation is performed at 5 kV for S4 and at 3 kV for S5 and S6.

**(X) Effect of hBN covering and carbonaceous contamination deposition on electron trajectories of 1L MoS$_2$:**

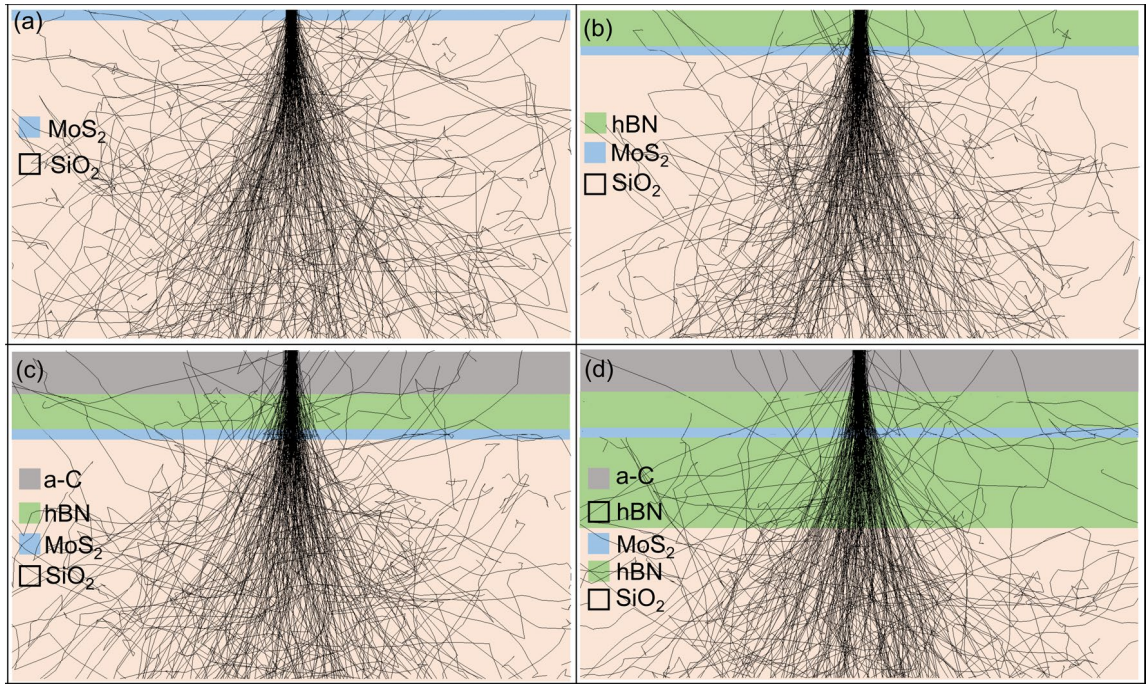

**Figure S8:** Monte Carlo simulations of electron trajectories for 1L MoS$_2$, hBN/MoS$_2$, amorphous carbon C/hBN/MoS$_2$, and C/hBN/MoS$_2$/hBN are shown in **(a)**, **(b)**, **(c)**, and **(d)** respectively. The thickness of top hBN, bottom hBN, and C used in the simulation is 8 nm, 20 nm, and 10 nm, respectively. All electron trajectory simulations were carried out using CASINO software at 3kV accelerating voltage.

To study the interaction of e-beam with ML MoS2, we performed Monte Carlo simulations of electron trajectories using CASINO software[41]. We consider four different cases: ML MoS$_2$, hBN (8 nm)/MoS$_2$, C (10 nm)/hBN (8 nm)/MoS$_2$, and C (10 nm)/hBN (8 nm)/MoS$_2$/hBN(20nm) (Fig. S8a-d). Electron energy of 3 kV is used in simulations, and trajectories of 200 electrons are shown. For ML MoS$_2$, the e-beam can easily interact with the sample to create defects. But in the case of the hBN /MoS$_2$ sample, the hBN layer restricts the interaction of carbonaceous contaminants with the ML MoS$_2$. However, the e-beam can still reach MoS$_2$ to create defects and cause electron doping. Further, increasing carbonaceous contaminants prevent defect formation and electron doping, by acting as a barrier layer to the e-beam.

**(XI) Electron irradiation of ML MoS$_2$ in SEM: changing electron dose by varying magnification:**

In order to study evolution of Raman peaks of ML MoS$_2$ in the dose range I (0 to 2$\xi$), we need to achieve lower electron doses ( <$10^4$ e$^-$ nm$^{-2}$ ). Lower electron dose can be achieved by irradiating MoS$_2$ MLs at different magnifications while keeping the beam current same as previous measurements. At 1000x magnification, the sample receives electron dose of $\rho$ (1.0486×10$^3$ e$^-$ nm$^{-2}$). Both E' and A$_1$' redshift with increased fullwidth half maximum attributed to defect formation and electron doping. The result is consistent with previous measurement done in dose range I, where electron dose is varied by changing scanning time.

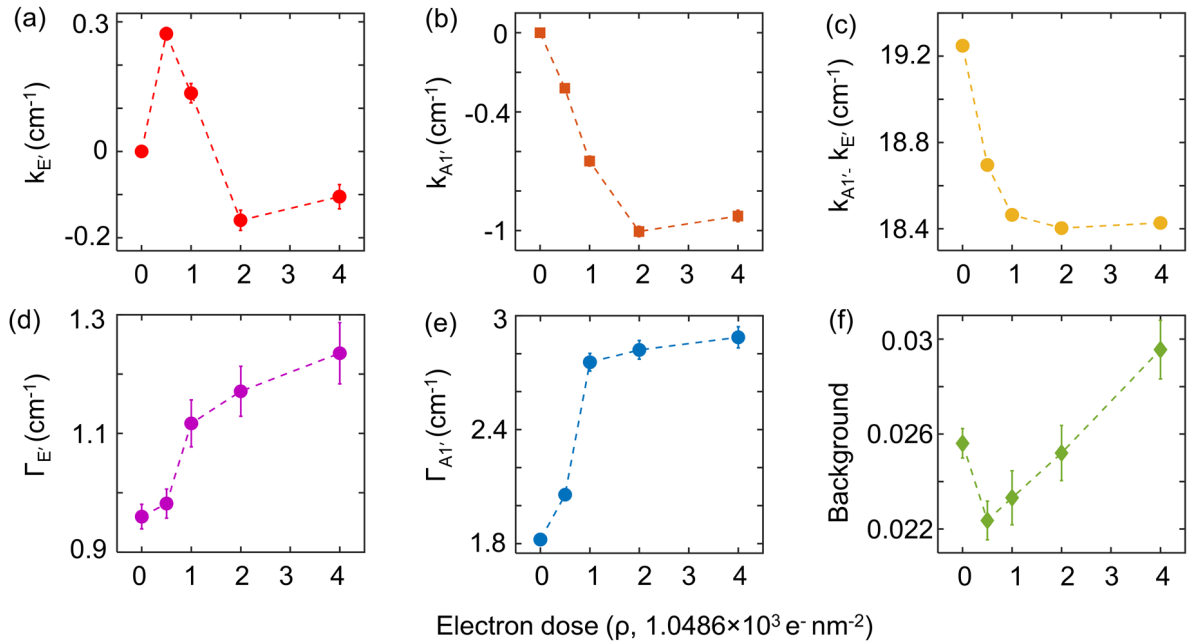

**Figure S9:** Variation of positions of E' and $A_1'$ peaks with electron dose is plotted in **(a)** and **(b)**, respectively. **(c)** The evolution of separation between E' and $A_1'$ peaks ($k_{A_1'} - k_{E'}$) with electron dose. Full width at half maximum ($\Gamma$) of E' and A' peaks with electron dose are shown in **(d)** and **(e)**, respectively. **(f)** The evolution of background of Raman spectra with electron dose is shown.

**(XII) Schematics showing contribution of defects and doping processes in E' and $A_1'$ peak changes with electron dose:**

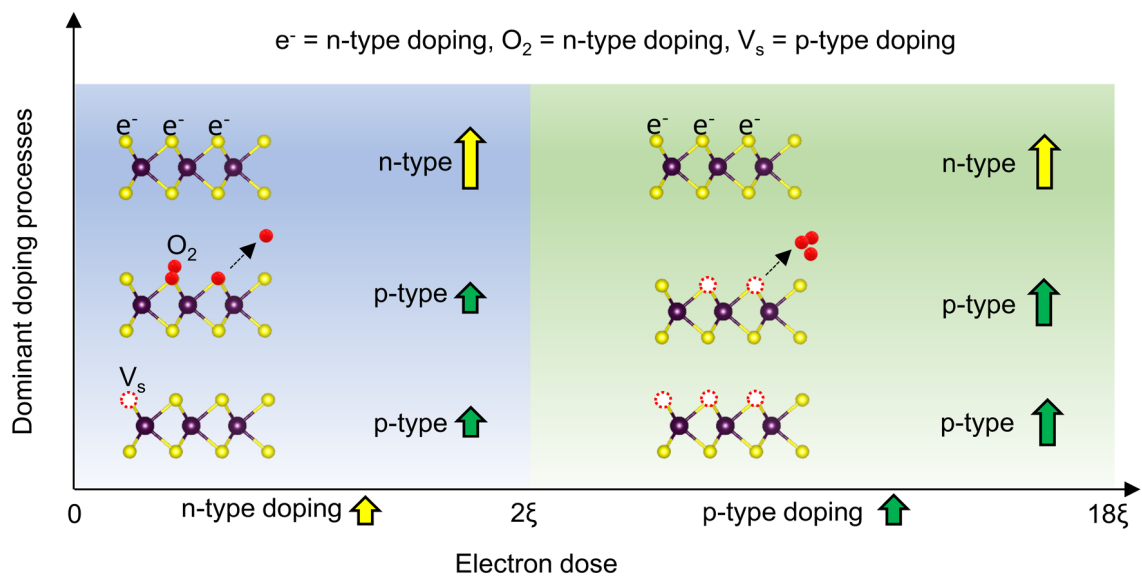

**Figure S10:** Schematics showing different doping processes during e-beam irradiation of TMDs. For dose range I (0-2$\xi$), n-type doping due to e-beam dominates over p-type doping effect of oxygen removal and defect formation. However, for dose range II (2$\xi$ - 18 $\xi$), p-type doping due to defect formation and oxygen removal dominates over n-type doping of e-beam resulting in an overall p-type doping.

**(XIII) Deconvoluting low energy L PL peak and temperature dependence:**

To study the nature of low energy PL peak (L), we performed power and temperature dependent PL of electron irradiated bare, hBN covered, and encapsulated ML $MoS_2$. The PL spectra of electron irradiated bare and hBN covered $MoS_2$ fitted to four peak Gaussian function are shown in Fig. S11(a) and (b), respectively. The excitonic peak ($X^0$) is highly quenched. However, we observed trion peak ($X^-$) in electron irradiated samples attributed to n-type doping due to e-beam. The variation of PL intensity with laser power for electron irradiated bare and hBN covered samples are plotted in Fig. S11(d) and (e), respectively. The L peak is fitted to D1, D2, and D3 peaks. Previously, D1, D2, and D3 were attributed to excitons bound to different defect species with primary defect type being sulphur vacancies[42]. However, exact nature of bound excitons in ML $MoS_2$ is still an open question and requires further study. The intensity of D1, D2, and D3 peak show linear dependence with the laser power (Table S1). On the other hand, the intensity of D1, D2, and D3 peaks decrease with increasing temperature and are vanishingly small at room temperature (Fig. S11(c)). Thus, the reduction in L peak (D1, D2, D3) intensity with temperature signifies the origin as defect bound excitons. The non-saturation with power indicates the high density of defects. Further, the appearance of L peak in PL spectra of all irradiated samples (uncovered, hBN covered, and encapsulated ML $MoS_2$) confirms that L peak corresponds to excitons bound to sulphur vacancies (or other defects) and the interaction of carbonaceous contaminants with sample is negligible.

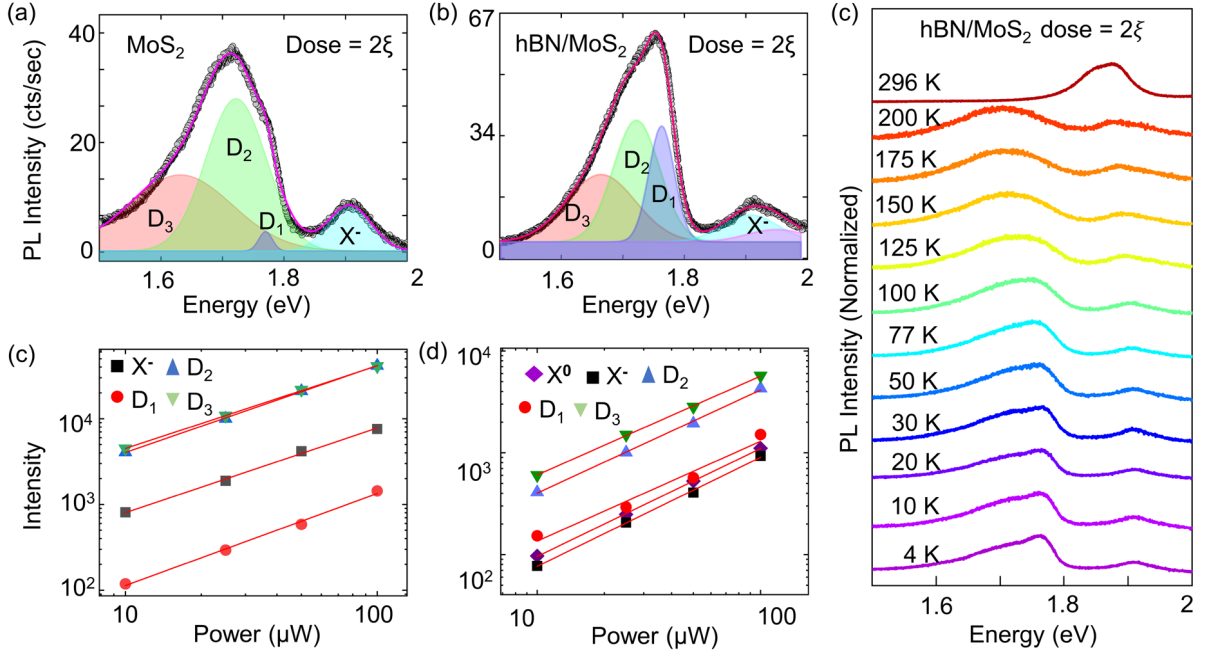

**Figure S11: Cryogenic photoluminescence (PL) of electron irradiated bare, hBN covered ML MoS$_2$. (a)**, **(b)** Defect (D1, D2, D3) and trion (X$^-$) PL peaks in electron irradiated bare and hBN covered ML MoS$_2$ obtained by Gaussian peak fitting. The laser power dependence of PL peaks for bare and hBN covered ML MoS$_2$ are shown in **(c)** and **(d)**, respectively. **(e)** Evolution of PL spectra for irradiated hBN covered ML MoS$_2$ with temperature. Here, $\xi = 1.3 \times 10^4$ e$^-$ nm$^2$.

| MoS$_2$ 2$\xi$ | | hBN/MoS$_2$ 2$\xi$ | |
|---|---|---|---|
| **Peak** | **Slope ($\alpha$)** | **Peak** | **Slope ($\alpha$)** |
| X$^-$ | 0.98737 ± 0.03444 | X$^-$ | 1.0674 ± 0.01126 |
| D$_1$ | 1.07265 ± 0.05133 | D$_1$ | 0.98788 ± 0.10191 |
| D$_2$ | 1.01335 ± 0.01128 | D$_2$ | 1.00637 ± 0.03407 |
| D$_3$ | 0.96356 ± 0.01087 | D$_3$ | 0.97171 ± 0.01234 |

**Table S1:** Power law coefficients for fitting performed in Figure S11.

## (XIV) Theoretical Framework: creation of the MLFF supermodel

The training of the forcefields began with the structural relaxation of pristine 4x4 supercell using DFT. The convergence criterion for relaxation was 0.002 eV/Å for the forces acting on atoms and the lattice parameters were allowed to change during the relaxation. A plane wave basis of 500eV was used to represent the wavefunctions and 3x3 k-point mesh was used to sample the Brillouin zone. Van Der Waals forces were simulated by using rev-vdW-DF2 functional[47].

The relaxed structure was then used in the molecular dynamics (MD) simulations for training the initial forcefield. A canonical ensemble was simulated by a Nose-Hoover thermostat[48,49]. Temperature was set to 500K with a timestep length of 1 fs and Nose-Hoover oscillation period of 50 fs. This higher temperature compared to experimental conditions was chosen to account for large deviations of atoms from the equilibrium. The forcefield was trained for 10000 ionic steps. The resulting forcefield of the pristine $MoS_2$ system was used as a starting point for learning different strains and defects in the system.

Beginning from the forcefield of the pristine system, continuation training was done using different strains. Four strains of -2%, -1%, 1%, and 2% were included in this study, from stretching (positive sign) to compressing (negative sign). Strains were applied by changing the lattice parameters of the relaxed 4x4 $MoS_2$ supercell by the indicated percentage, relaxing the structure using DFT with fixed lattice parameters and performing a MD run at a temperature of 300K corresponding to experimental conditions. This retraining of the pristine forcefield was done for 10000 ionic steps for each strain, yielding a set of four forcefields.

The effects of sulfur vacancies and oxygen defects were also included in the forcefields. The pristine structure of the 4x4 super cell was used as a template to create the defective systems. Several defects were initially considered (SI - XV), but finally only three defect types remained relevant: substitutional and adatom defects of oxygen along with the sulfur vacancy. The structures containing these three defect configurations were first relaxed with DFT and then a MD run was done similar to training of different strains beginning from the forcefield of the pristine system. This yielded three different forcefields with information on the influence of different defects.

We then combined the configurations obtained during the training of pristine and strained $MoS_2$ as well as the defective systems into a singular supermodel. We scanned this set of configurations to look for possible similar ones (similar to what is done during the on-the-fly training). Note that only 7 out of the 1242 configurations were removed during this scan, showing that training different set in parallel was indeed reasonable. This resulting set of 1235 configurations was then used to find the final MLFF model. Diagram showing the training workflow is available in Figure S12, along with benchmarks of the phonon dispersion curve and the impact of strain on phonon frequencies.

The calculated raw formation energies of defects using the final model were compared against DFT results [Table S2] and the differences in formation energy between the DFT and the MLFF were found to be negligible. A similar test was done by comparing the energy change of the system under biaxial strain [Table S3], showing that with increased strain the deviation of MLFF energies from DFT energies remained small.

The MLFF model was finally used to calculate the vibrational spectra of defective $MoS_2$. 10x10 $MoS_2$ supercells were adopted, and one to five defects were inserted in the supercell corresponding to 1%-5% defect concentration. The defects were allowed to be on either side of the monolayer and the minimum distance between them was set to 10 Å. The locations were the same for each defect type with the same concentration. The structures were relaxed using the supermodel while keeping the lattice parameters fixed matching those of the pristine $MoS_2$. The force constants for obtaining vibrational eigenmodes were calculated using finite-difference approach.

## (XV) Theoretical Framework: other considered defects

The e-beam itself could facilitate vacancy defect formation. On the other hand, oxygen atoms or $O_2$ molecules could be adsorbed onto the vacancies or to the pristine surface. In total, thirteen defect configurations were investigated. This included three types of defects: sulphur vacancies, single oxygen atoms and $O_2$ molecules.

As the periodicity of the $MoS_2$ pristine structure allows for only one unique case of a sulphur vacancy, only one sulphur vacancy configuration was then included in the study. The $O_2$ molecule was placed on three different sites on the pristine monolayer: on top of Mo and S atoms, as well as on to the empty site in the middle of hexagons. The sulphur vacancy position was also used as a site for inserting the $O_2$ molecule. The $O_2$ molecule was placed on each of the four sites both horizontally and vertically, making up eight defect configurations in total. Single oxygen atoms were placed in similar locations adding four more configurations.

Out of the initial oxygen defect configurations only four were found to be stable. During ionic relaxation using DFT, the vertical $O_2$ configurations along with the horizontal configuration of $O_2$ on hexagonal site detached from the surface. The following MD steps used for training the forcefields detached four defect configurations: oxygen adatoms on hexagonal- and Mo sites, horizontal $O_2$ on hexagonal site, and vertical $O_2$ on top of the S site.

After DFT and MD, two oxygen atom defects remained bonded to the surface: O adatom on top of S and substitutional O on S site. Two $O_2$ molecule configurations also remained bonded to the surface: horizontal configurations on sulphur vacancy and on top of sulphur. Counting the stable oxygen defect configurations and the sulphur vacancy, five defect configurations then were left to consider out of the initial thirteen.

However, in both stable $O_2$ configurations, the molecule was split into two oxygen atoms that remained within proximity to one another. The uniqueness of these configurations should be questioned; can these $O_2$ configurations be considered as just two separate oxygen defects?

In the case of $O_2$ on sulphur vacancy, two neighbouring defects of substitutional oxygen and oxygen adatom were formed. The formation energy of this defect configuration is only 0.02 eV higher than the formation energies of separate adatom and substitutional defects (Table S2). From the small difference it can be concluded that the interaction between these neighbouring defects is very weak. $O_2$ defect can then be discarded as unique case and can be regarded just as separate O adatom and O substitutional defects.

$O_2$ defect on top of a sulphur atom consists of two oxygen adatom defects attached to the same sulphur atom. The $O_2$ configuration formation energy is 0.466 eV higher than the formation energies of two O adatom defects (Table S2). Thus, under experimental conditions, this $O_2$ configuration is then very likely to dissociate into two O adatom defects on different S sites and is thus not considered further. Finally, only three defect configurations were used to train the forcefields: sulphur vacancy, oxygen adatom and oxygen substitution defect on S site.

| Defect: | $E_f$ (eV) | Separate defect $E_f$ (eV) | $\Delta E_f$ (eV) |
|---|---|---|---|
| $O_2$ top of S site | −7.66 | −8.13 (two O adatoms) | 0.47 |
| $O_2$ on vacancy | −6.05 | −6.08 (substitutional O and O adatom) | 0.02 |

**Table S2**: Formation energies of two additional $O_2$ defect configurations calculated using MLFF and comparison to the (sum of) formation energies after dissociation. The formation energies are defined as: $E_f = E_{defective} - E_{pristine}$, which corresponds to S and O chemical potentials taken from neutral atom.

## (XVI) Benchmarking of MLFF

| Defect: | DFT $E_f$ (eV) | MLFF $E_f$ (eV) | $\Delta E_f$ (eV) |
|---|---|---|---|
| O substitutional | -2.01 | -2.01 | 0.001 |
| O adatom | -4.07 | -4.06 | 0.004 |
| S vacancy | 5.58 | 5.58 | -0.008 |

**Table S3:** Comparison of defect formation energies calculated using DFT and MLFF. The formation energies are defined as: $E_f = E_{defective} - E_{pristine}$, which corresponds to S and O chemical potentials taken from neutral atom.

| Strain: | DFT $E_f$ (eV) | MLFF $E_f$ (eV) | $\Delta E$ (eV) |
|---|---|---|---|
| -2% | 0.61 | 0.59 | -0.018 |
| -1% | 0.15 | 0.14 | -0.008 |
| 1% | 0.15 | 0.16 | 0.010 |
| 2% | 0.57 | 0.59 | 0.019 |

**Table S4:** Comparison of "strain energy" of pristine MoS$_2$, defined as: $E_s = E_{strained} - E_{pristine}$, calculated using DFT and MLFF and for different values of biaxial strain.

In quantum chemistry, "gold standard" refers to method that gives the best, most accurate results and ideally gives results that are comparable to experimental accuracy (~1 kcal/mol for energies). This method is usually CCSD(T). Since DFT is not as accurate as CCSD(T) and since the calculated vibrational frequencies clearly differ from experiments, we cannot claim overall "gold standard" accuracy.

We have assessed the reliability of the MLFF vs DFT in predicting vibrational properties. MLFF results were found comparable to DFT results to an accuracy of about 1 cm$^{-1}$ in absolute frequencies and likely much less in relative changes, which would be comparable to experimental accuracy.

But how good are the DFT results vs experiments? In the pristine system, the Raman mode frequencies differ by few cm$^{-1}$, but when it comes to the role of defects, this is a very difficult question. To assess the reliability of the calculations, we would need to compare to more accurate calculations or known experimental results. More accurate calculations are computationally intractable, and in the case of DFT the results cannot be systematically improved by adding higher-order terms and thus computationally heavier methods are not guaranteed to always yield more accurate results. Moreover, there are no extensive experimental references for Raman peak shifts upon introduction of known density of known defects in known positions to an otherwise pristine host. This makes the comparison challenging.

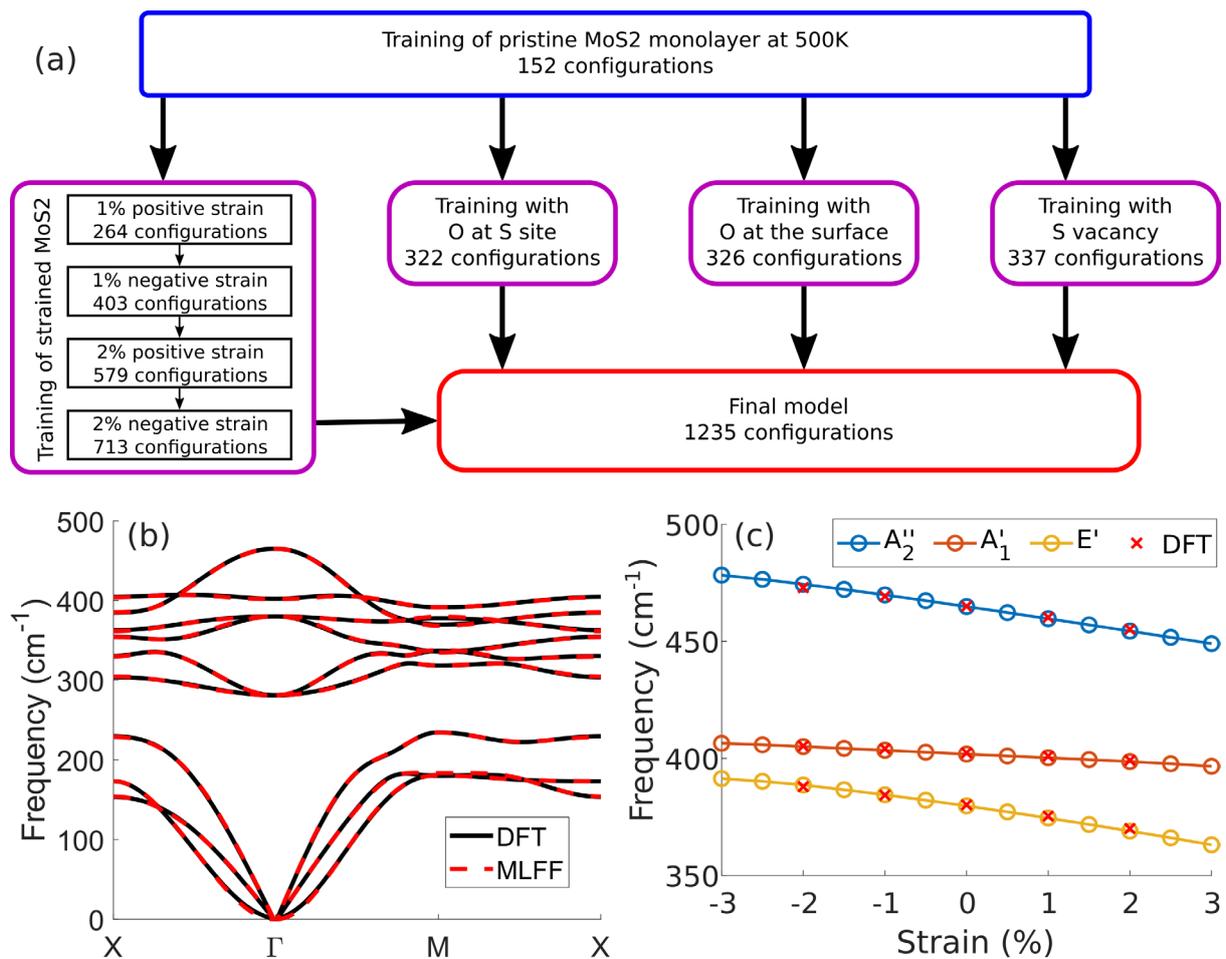

**Figure S12:** (a) Diagram of the MLFF model training. (b) Phonon dispersion curves using the final MLFF model and DFT. (c) Phonon frequency as a function of strain using the MLFF model. Red crosses show the DFT results for comparison.

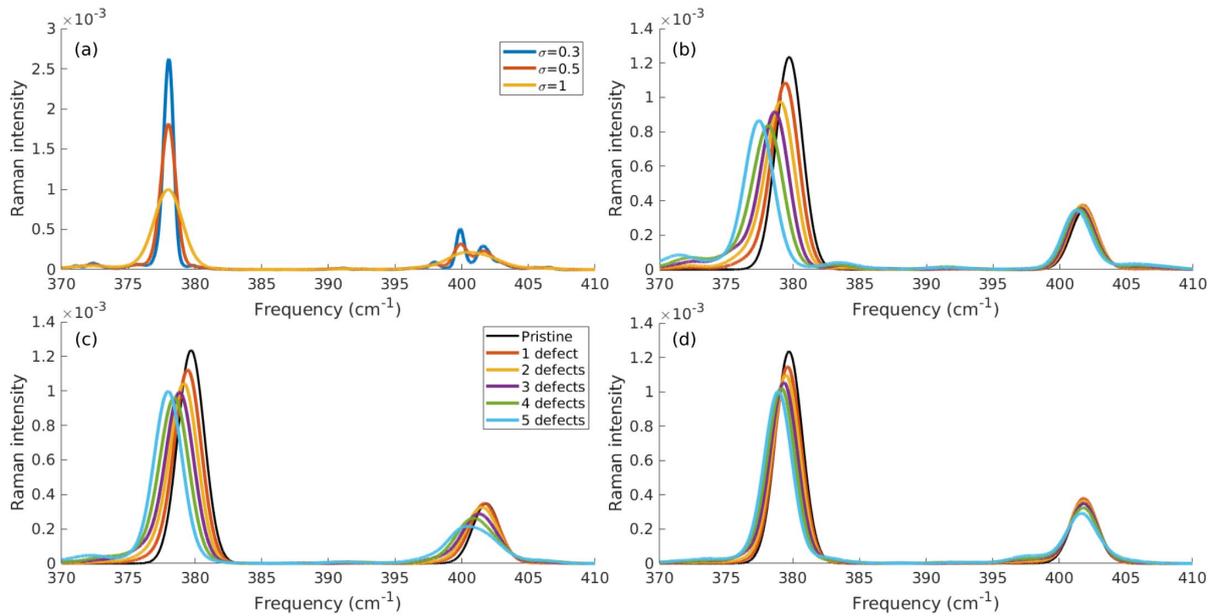

**Figure S13:** (a) Raman spectra of 10x10 supercell with 5 oxygens at sulphur sites using different broadening. (b-d) Raman spectra for increasing amount of sulphur vacancies, oxygen at sulphur sites, and oxygen adatoms at the surface, respectively.

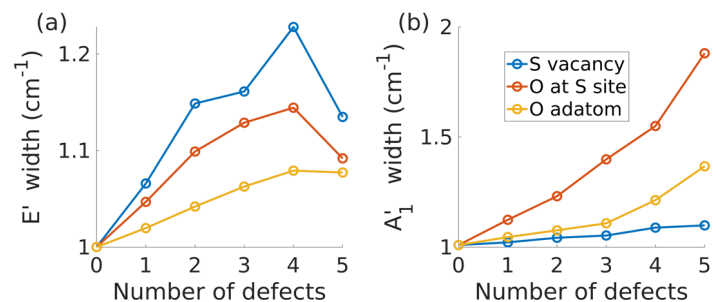

**Figure S14:** Full width at half maximum (FWHM) for (a) the E' and (b) the $A_1'$ modes as a function of the number of defects. Values are obtained using a Gaussian fit on the Raman spectra shown in Fig. S13.

**(XVII) Monte-Carlo simulation of backscattered electrons**

We used Monte-Carlo modelling, as shown in Fig. S15 (Also discussed in Supporting Information X). The number of back-scattered electrons compared to the incident electrons is low. Only 10% of total incident electrons get back-scattered for irradiation at 5 kV electron accelerating voltage (back-scattering coefficient = 0.1). Also, the energy of back-scattered electrons is small, compared to incident electrons, and thus less damage is expected. As shown in Figure S15, backscattered electrons can increase the damage area beyond the direct incidence beam. Sputtered silicon atom can cause damage, but the probability of sputtering is small at ultralow electron accelerating voltage.

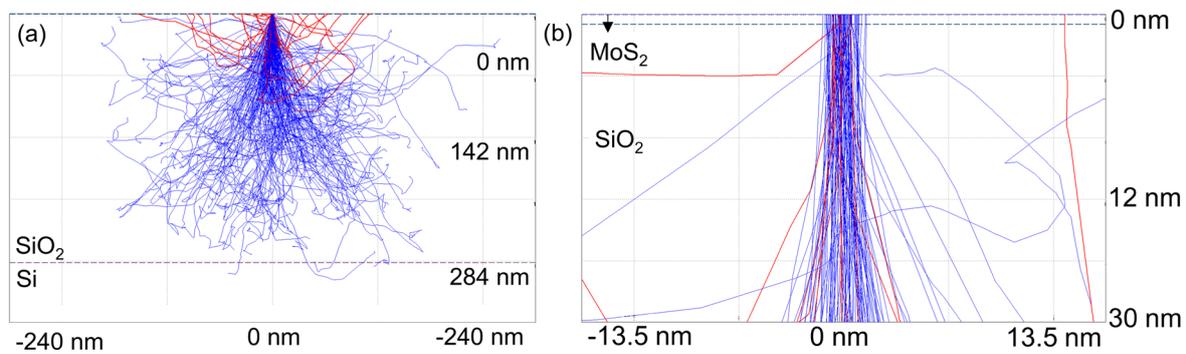

**Figure S15.** (a) Monte-Carlo simulation of electron trajectories of monolayer $MoS_2$ on $SiO_2$ (285nm)/Si at 5 kV electron accelerating voltage. The backscattered electrons are shown in red. Only 10% of electrons gets backscattered (coefficient = 0.1). (b) zoomed in version of (a) to illustrate interactions near the monolayer $MoS_2$.

**(XVIII) Damage mechanisms at ultralow accelerating voltage electron irradiation**

For irradiation above threshold electron accelerating voltage (80 kV for S atom removal in monolayer $MoS_2$), the defects are formed by knock-on damage mechanism [7,14,15]. However, defects can still be formed below knock-on threshold electron accelerating voltage by additional energy channels in the system [19, 20, 21]. For ionization and radiolysis processes, the energy transferred to the medium increases as accelerating voltage decreases, scaling as 1/E. The dependence is not expected as 1/E, but levelling off and eventually decreasing at low acceleration voltages of about 1 kV [37]. Further, recent studies have discussed the reduction of displacement threshold energy of sputtering atoms due to localization of electronic excitations to emerging defect sites [19].

We note, at ultralow acceleration voltage (1-10 kV), the displacement cross-section is still non-zero and defect formation can happen. Adsorbed oxygen atoms may result in lowering the barrier even further for damage, by increasing localization of electronic excitations. We hope our work will motivate researchers to perform further studies for understanding the defect formation mechanism at ultralow electron accelerating voltage. Further, we are not aware of time-dependent DFT simulations directly showing sputtering by ionization, and they would be challenging due to the low probability of such events.

We also note that we have proven experimentally that carbonaceous contaminants deposited during irradiation do not participate in defect formation process, resolving a key question [20, 21]. This is a major contribution to field of defect creation using ultralow accelerating voltages.

**(XIX) Raman and PL spectra in log scale**

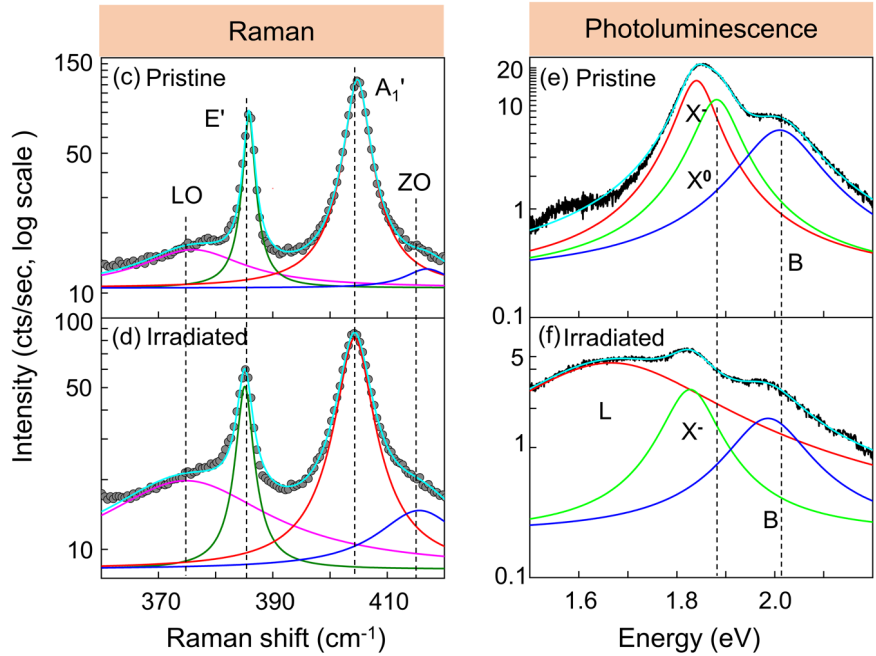

**Figure S16.** Plot of intensity of Raman (c, d) and PL peaks (e, f) in the logarithmic scale for Figure 1.

**(XX) Comparison of 3 kV and 5 kV irradiation room temperature PL data**

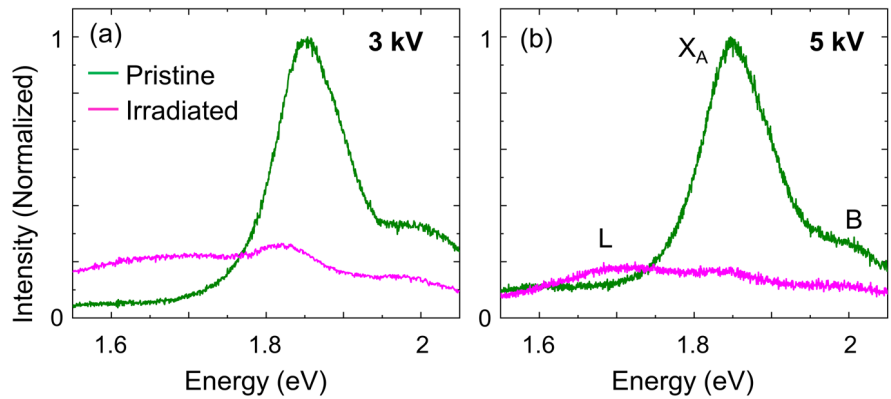

**Figure S17.** Comparison of PL intensity quenching of irradiated bare monolayer MoS$_2$ (dose = 2ξ) for 3 kV (a) and 5 kV (b) accelerating voltage. 5 kV electron irradiation shows more reduction in A-exciton intensity compared to 3 kV irradiation. The spectra are normalized to A exciton intensity of pristine sample. These spectra are taken at room temperature.

**(XXI) Statistical results of peak positions and FWHM of 3 kV and 5 kV irradiated samples**

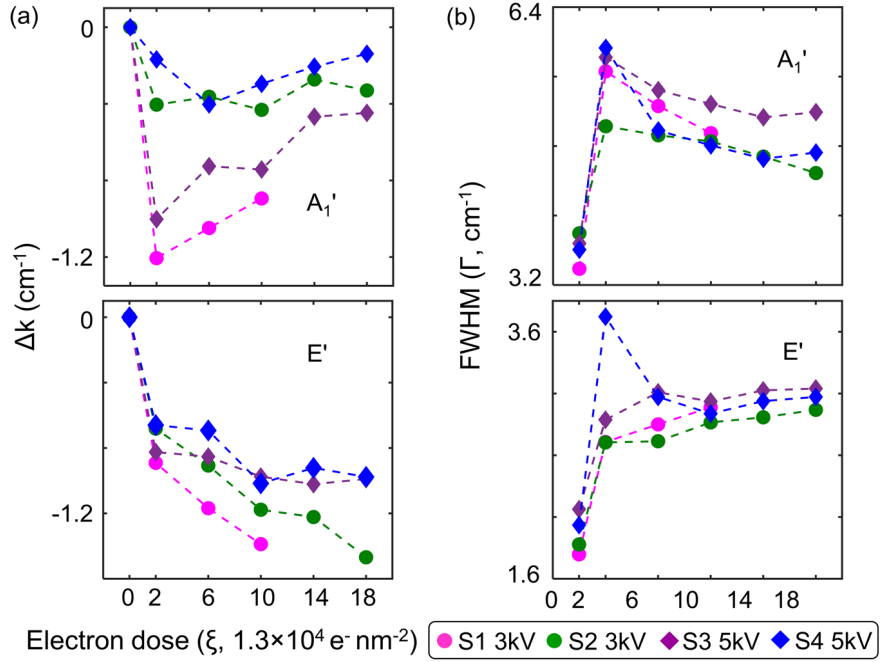

**Figure S18.** Evolution of position and FWHM of $A'_1$ and $E'$ Raman peaks of four bare monolayer $MoS_2$ with electron dose for 3 kV and 5 kV electron accelerating voltages.

**(XXII) Raman mapping of E' and $A_1'$ modes of electron irradiated monolayer $MoS_2$**

We have performed Raman mapping of irradiated monolayer $MoS_2$ in a Renishaw system using a 532 nm laser, 50x objective, and 2400 lines/mm grating. The obtained spectral array was analysed using MATLAB. Mapping images of frequencies and FWHM of the fitted E' and $A_1'$ Raman modes are shown in the below figure. Spatial variation of frequencies and FWHM of E' and $A_1'$ Raman modes (for different dose) matches well with trends presented for point spectra in Figure 2 (main text).

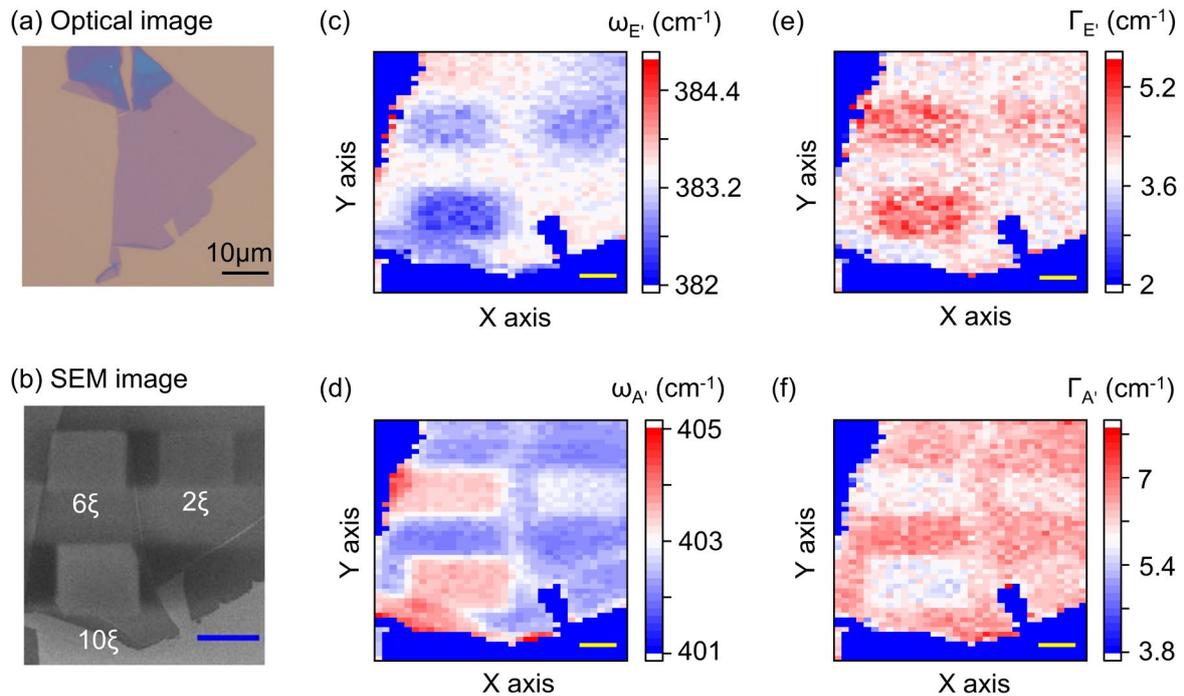

**Figure S19.** (a) Optical image of pristine monolayer MoS$_2$. The SEM image after electron irradiation at 3 kV is shown in (b), showing regions irradiated with different dose. Here, ξ = 1.3×10$^4$ e$^-$ nm$^{-2}$. Spatial maps of frequencies of E' and A$_1$' Raman modes are plotted in (c) and (d), respectively. Spatial maps of FWHM of E' and A$_1$' Raman modes are shown in (e) and (f), respectively. Scale bars in (b-f) is 5μm.